\documentstyle{mn}
\input{epsf}
\def\deg{\hbox{$^\circ$}}

\title[Superlarge-Scale Structure]
{{Superlarge-Scale Structure in N-body simulations}}

\author[Doroshkevich et al.]  {A.G. Doroshkevich$^{1,2}$,
       V. M\"uller$^3$, J. Retzlaff$^3$, and V. Turchaninov$^2$\\
       $^1$Theoretical Astrophysics Center, Juliane Maries
       Vej 30, DK-2100 Copenhagen \O, Denmark \\
       $^2$ Keldysh Institute of
       Applied Mathematics, Russian Academy of Sciences,
	125047 Moscow, Russia\\
       $^3$ Astrophysikalisches Institute Potsdam,
	An der Sternwarte 16, Potsdam, D-14482 Germany}
\date{Accepted ..., Received 1998  ...; in original form 1998  ...}

\begin{document}
\maketitle

\begin{abstract}
The simulated matter distribution on large scales is studied using
core-sampling, cluster analysis, inertia tensor analysis, and minimal
spanning tree techniques.  Seven simulations in large boxes for five
cosmological models with COBE normalized CDM-like power spectra are
studied.  The wall-like Super Large Scale Structure with parameters
similar to the observed one is found for the OCDM and $\Lambda$CDM models
with $\Omega_m h = 0.3 ~\&~ 0.245$.  In these simulations, the rich
structure elements with a typical value for the largest extension of
$\sim$$(30 - 50)h^{-1}$Mpc incorporate $\sim$40\% of matter with
overdensity of about 10 above the mean.  These rich elements are formed
due to the anisotropic nonlinear compression of sheets with original size
of $\sim$$(15 - 25)h^{-1}$Mpc.  They surround low-density regions with a
typical diameter $\sim$$(50 - 70)h^{-1}$Mpc.

The statistical characteristics of these structures are found to be
approximately consistent with observations and theoretical expectations.
The cosmological models with higher matter density $\Omega_m=1$ in CDM
with Harrison-Zeldovich or tilted power spectra cannot reproduce the
characteristics of the observed galaxy distribution due to the very
strong disruption of the rich structure elements.  Another model with a
broken scale invariant initial power spectrum (BCDM) shows not enough
matter concentration in the rich structure elements.
\end{abstract}


\section{Introduction}

The phenomenon of Super Large Scale Structure (SLSS) was first observed as
rare peculiarities in the visible galaxy distribution, with extreme
parameters.  Examples of this include the Great Void (Kirshner et al.
1983), the Great Attractor (Dressler et al.  1987), the Great Wall (de
Lapparent, Geller, \& Huchra 1988; Ramella, Geller, \& Huchra 1992), and the
Pisces-Perseus supercluster (Giovanelli \& Haynes 1993).  Several nearby
superclusters of galaxies were described by Oort (1983a, b).  SLSS was also
found in deep pencil beam redshift surveys (Broadhurst et al.  1990;
Willmer et al.  1994; Buryak et al.  1994; Bellanger \&~ de Lapparent 1995;
Cohen et al.  1996) as rich galaxy clumps with typical separations in the
range of $(60 - 120)h^{-1}$Mpc ($h$ is the Hubble parameter in units of
100 km s$^{-1}$ Mpc$^{-1}$).

Recently the analyses of rich galaxy surveys with an effective depth
$\sim$$(200 - 400)h^{-1}$Mpc, such as the Durham/UKST Galaxy Redshift
Survey (Ratcliffe et al.  1996) and the Las Campanas Redshift Survey
(LCRS) (Shectman et al.  1996), have established the existence of the
wall-like SLSS as typical phenomenon in the visible galaxy distribution
incorporating $\sim$$(40 - 50)$\% of galaxies (Doroshkevich et al.  1996,
hereafter LCRS1; Doroshkevich et al.  1997b, hereafter LCRS2; 1998a;
1999).  The wall-like SLSS consists of structure elements with a typical
diameter $\sim(30 - 50)h^{-1}$Mpc surrounding low-density regions (LDR)
with a similar typical diameter $D_{LDR}\sim (50 -- 70)h^{-1}$Mpc.
Within the wall-like SLSS elements the observed galaxy distribution is
also inhomogeneous (see, e.g., Fig.  5 of Ramella et al.  1992), and
galaxies are concentrated in high density clumps and filaments.

In LDR the galaxies are found to be concentrated within a random network
of filaments.  In distinction to the typical wall-like superclusters, the
galaxy distribution in LDR is predominantly 1-dimensional with a mean
separation of filaments $\sim$$(10 - 15)h^{-1}$Mpc (LCRS1), and we call
this network Large Scale Structure (LSS).  The LSS incorporates also
$\sim$50\% of galaxies and is clearly seen in many redshift catalogues of
galaxies (see, e.g., de Lapparent, Geller \& Huchra 1988).  These results
extend the range of investigated scales in the galaxy distribution up to
$\sim 100h^{-1}$Mpc.

While the LSS was predicted by the nonlinear theory of gravitational
instability (Zel'dovich 1970), and it was reproduced in simulations before
its discovery in observations (see, for reference, Shandarin \& Zel'dovich
1989), the observation of the rich and typically wall-like SLSS was quite
unexpected.  Thus, in simulations the representative SLSS was found only
recently for a CDM model with low density and a cosmological constant
($\Lambda$CDM, Cole et al.  1997).  This simulation demonstrates that for
suitable parameters of cosmological models the formation of the wall-like
SLSS is compatible with the standard CDM power spectrum of Gaussian initial
perturbations.

The formation and evolution of structure on all scales are described by an
approximate theoretical model (Demia\'nski \& Doroshkevich 1998a, b) based
on Zel'dovich nonlinear theory of gravitational instability (Zel'dovich
1970; 1978; Shandarin \& Zel'dovich 1989).  The model shows that the SLSS
formation can be related to matter infall into the large wells of the
gravitational potential of the initial perturbations.  This model connects
the structure parameters with the main parameters of the underlying
cosmological scenario and the initial power spectrum.  It shows that the
impact of large scale perturbations is important throughout all
evolutionary stages.  In particular, the influence of these perturbations
modulates the merging of smaller structure elements promoting the
evolution within the SLSS elements and depressing it in LDR.

The simulations are able to take into account this interaction throughout
all evolutionary stages and, therefore, are the most suitable way to study
the LSS and SLSS properties and evolution.  This implies, however, that the
simulations need to be performed in very large boxes both to provide us
with reasonable statistics of the walls and, of particular importance, to
correctly describe the large scale part of the initial power spectrum of
perturbations, responsible for the wall formation and the mutual
interaction of small and large scale perturbations.  In practice, this
means that we need extreme parameters in the simulations.  On the other
hand, in order to compare various cosmological models, a broad set of
simulations has to be prepared.

Here results are presented from the analysis of simulations using five
cosmological models.  It is shown that the models with high matter density,
$\Omega_m\sim 1$, cannot reproduce adequately the observed properties of
the large scale matter distribution.  The models with lower matter density,
in particular the OCDM and $\Lambda$CDM models, seem to be more promising,
since they can reproduce the general observed characteristics of both the
LSS and the SLSS.  In these models the rich structure elements, formed by
a nonlinear matter compression, contain a significant matter fraction,
$f_{rse}\sim 0.4 - 0.5$, which can be easily discriminated.  A more
detailed investigation of the nonlinear matter evolution on large scales
is a further goal of this paper.

The small scale matter clustering resulting in the destruction of
structure elements restricts the class of cosmological models which are
capable of reproducing the observed LSS and SLSS.  The instability of a
sheetlike matter distribution similar to the observed and simulated SLSS
was considered (in the linear approximation) by Doroshkevich (1980) and
Vishniac (1983), and it was recently simulated by Valinia et al.  (1997).

Following our previous papers (LCRS1; LCRS2; Doroshkevich et al. 1997) we
concentrate on the geometrical properties of the matter distribution, in
particular, to the proper sizes and spatial distribution of filaments and
wall-like structure elements.  The popular correlation analysis is not so
useful at scales $>10h^{-1}$Mpc discussed below, and other techniques
provide us with more essential results.  We employ the core-sampling
approach introduced by Buryak et al.  (1994), the standard cluster analysis
supplemented by the inertia tensor technique (Vishniac 1986; Babul \&
Starkman 1992), the analysis of the variations of number of clusters vs.
linking length (NCLL method, Doroshkevich et al.  1997b), and the minimal
spanning tree (MST) technique (Barrow et al.  1985; van~de~Weygaert 1991).
These methods were utilized recently for the investigation of structures
in the LCRS (cp.  LCRS1 and LCRS2) and Durham/UKST redshift surveys
(Doroshkevich et al. 1999).  These results shall be used for the
comparison with the structure parameters derived from simulations.  The
different methods are complementary to each other, and, thus, they allow
us to characterize the simulated matter distribution in different
important aspects.

The observed distribution of galaxies and the simulated distribution of
the DM cannot be identical as the galaxy formation is sensitive to
additional factors (e.g., to the reheating) and, moreover, galaxies,
probably, mark only the highest peaks of density perturbations.  This
means that some parameters of DM structure elements such as their
overdensity and proper sizes can differ from that found in observational
catalogues.  The comparative analysis performed for one simulation
(Doroshkevich et al.  1998a) confirms, that in some respect the spatial
distributions of DM and `galaxies' are different.  A more detailed
comparison of the observed and simulated matter distribution implies an
identification of `galaxies' in the simulated DM distribution.  This means
that a certain bias model needs to be utilized (see, e.g., discussions in
Sahni \& Coles 1995, and Cole et al.  1998).  Both problems are, however,
equally important, and the distributions of both the galaxies and the DM
must be studied.

This paper is organized as follows.  The simulations and the analysis
techniques utilized are shortly described in Secs.  2 and 3.  In Sec.  4
the general characteristics of the considered simulations are discussed
that allow us to roughly discriminate the cosmological models and to select
the most realistic ones for a more detailed investigation.  Our main
results are presented in Secs.  5 and 6.  Sec.  7 is devoted to the
comparison with theoretical expectations of the DM distribution, and in
Sec.  8 we discuss mock galaxy catalogues using some simple bias models.
The conclusion and a discussion can be found in Sec.  9.

\section{Simulations}

We used five cosmological models as a basis for our analysis.  The COBE
normalized SCDM model is taken as a reference model despite its
difficulties in explaining already standard measures of galaxy clustering as
the power spectrum and the correlation function of galaxies and galaxy
clusters (cp.\ e.g.\ Ostriker, 1993).  Alternative models with
$\Omega_m=1$ include modifications of the primordial power spectrum, in
particular by introducing a tilt $\propto k^{0.9}$ of the power spectrum
(TCDM), or a break at a certain scale, (BCDM).  Both are
inflation motivated, using either an exponential inflation potential
(Lucchin \& Matarrese 1985), or a double inflation scenario (Gottl\"ober,
M\"uller, \& Starobinsky 1991).  The BCDM is specified by two parameters,
the location of the break at $k_{\rm{break}}^{-1}=1.5h^{-1}$Mpc, and the
relative power on both sides of the break, $\Delta=3$.  These parameters
were originally chosen to obtain optimal linear fits to the
various large-scale structure observations (Gottl\"ober, M\"ucket, \&
Starobinsky 1994), and later tested against $N$-body simulations (Amendola
et al.  1995; Kates et al.  1995; Ghigna et al.  1996; Retzlaff et al.
1998).  Both TCDM and BCDM models seemed to be promising since they have
reduced power at galactic scales with respect to the COBE normalized SCDM
model.

\begin{figure}
\centering
\epsfxsize=9 cm
\epsfbox{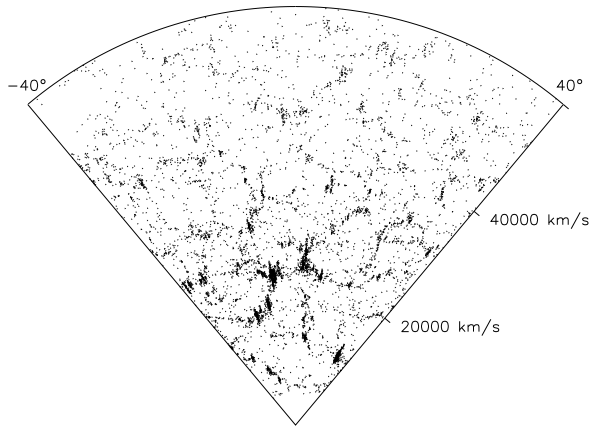}
\vspace{0.2cm}
\caption{Simulated point distribution in redshift
space for $\Lambda$CDM model at redshift z=0}
\end{figure}

\begin{table}
\caption{Parameters of simulations.}
\begin{tabular}{cccccccc}
model &$\Omega_m$&$h$&$L_{box}$&$N_{p}$&$N_{cell}$&$
\sigma_8$&$\sigma_{vel}$\cr
SCDM1        &1~~~~~&0.5& 500 & $300^3$& $600^3$  & 1.05& 1157\cr
SCDM2        &1~~~~~&0.5& 400 & $256^3$& $512^3$  & 1.12& 1127\cr
TCDM~        &1~~~~~&0.5& 500 & $300^3$& $600^3$  & 1.25& 1293\cr
BCDM~        &1~~~~~&0.5& 500 & $300^3$& $600^3$  & 0.60&~~714\cr
OCDM~        &0.5~~ &0.6& 500 & $300^3$& $600^3$  & 0.74&~~550\cr
$\Lambda$CDM1&0.35  &0.7& 500 & $300^3$& $600^3$  & 1.12&~~913\cr
$\Lambda$CDM2&0.35  &0.7& 400 & $256^3$& $512^3$  & 1.30&~~873\cr
\end{tabular}
$\Omega_m$ is the matter density, $h$ is the dimensionless Hubble
constant, $L_{box}$ is the box size in $h^{-1}$Mpc, $N_{p}$ and $N_{cell}$
are the particle and cell numbers, $\sigma_8$ is the mass variance for the
linear spectrum at the scale $8h^{-1}$Mpc, $\sigma_{vel}$ is the variance
of linear particle velocity in km/s.
\end{table}

Furthermore, two models are studied which are based on the wide range of
observations pointing to a lower matter density in the universe.  First,
we study an OCDM model with $\Omega_m=0.5$ violating the inflationary
paradigm of a spatially flat universe.  Second, we take a model with a
cosmological constant, which has $\Omega_m=0.35$ and a vacuum energy
leading again to a spatially flat universe.  This $\Lambda$CDM model bears
some advantage in alleviating the tight age constraint of the universe.
For all models the standard para\-meterization of the CDM transfer
function of Bardeen et al.  (1986, hereafter BBKS) was used.
In Table 1 the main
parameters of simulations are listed, including the matter density
$\Omega_m$, the dimensionless Hubble constant $h$, the box size
$L_{box}$, the particle number $N_{p}$ and the cell number $N_{cell}$.
Two models (SCDM and $\Lambda$CDM) were simulated with different
resolutions and with slightly different initial amplitudes.

The power spectra were normalized according to the two year COBE
measurement following Bunn et al.  (1995), (SCDM1, TCDM, BCDM, and
$\Lambda$CDM1), or to the four year data according to the description of
Bunn and White (1997), (SCDM2, OCDM, $\Lambda$CDM2), in both cases assuming
pure adiabatic perturbations and a baryon content of $\Omega_b h^2 =
2\cdot 10^{-2}$ as predicted by big bang nucleosynthesis (see, e.g.,
Schramm 1998).  Our later
discussion will show that the smaller amplitude of the four year
normalization, or a contribution of gravitational waves and/or other
inhomogeneities to the COBE signal do not influence significantly the main
conclusions.  The amplitude of perturbations is characterized by the mass
variance for the linear spectrum at the scale $8h^{-1}$Mpc, $\sigma_8$, and
the 3-dimensional velocity dispersion, $\sigma_{vel}$, gained from all dark
matter particles in the simulation.

The simulations were run in boxes of comoving size $L_{box}=500h^{-1}$Mpc
and $L_{box}=400h^{-1}$Mpc, respectively, to provide a good statistics of
perturbations in the range of wave numbers $k^{-1} \sim (10 -
30)h^{-1}$Mpc responsible for the SLSS formation and, so, to improve the
statistical characteristics of SLSS elements and the description of the
interaction of small and large scale perturbations.  For such boxes the
formation of the majority of structure elements is described by higher
harmonics of the primordial density waves, with $l \geq 8 - 10$.  We use
the PM code, described in more detail in Kates et al.  (1995) and
Retzlaff et al.  (1998), with $N_{\rm{p}} = 300^3 \,\mbox{or}\, 256^3$
particles in $N_{cell}=600^3 \,\mbox{or}\, 512^3$ grid cells,
respectively.  These parameters provide a resolution $\sim 0.9h^{-1}$Mpc
and the mass resolution $\sim$$(1 - 3)\cdot 10^{11}M_\odot$.

Most statistics can be calculated only for subsamples of the huge data
sets.  Therefore, we mostly used slices of thickness 50$h^{-1}$Mpc of the
simulation box, that are about 10\% of the complete volume.  Even this
volume provides us with a reasonable representation of the SLSS elements.
This high stability of structure parameters is a direct consequence of the
large box size used. To test the reliability, we repeated the analysis for
different slices and constructed some statistics for the full volume taking
the $\Lambda$CDM1 model. The main difference between the analysis of the
full sample and slices is a variations of the mean velocity dispersion of
clusters by  $5 - 7$ percent. Similar differences are characteristic
of the different realizations of the same cosmological model (SCDM1 vs.
SCDM2, $\Lambda$CDM1 vs. $\Lambda$CDM2). Below we give the basic results
for the larger simulations only, they are denoted as SCDM and $\Lambda$CDM.

The comparison of simulated and observed parameters of the SLSS has to be
done in the redshift space while the theoretical predictions are usually
made in the comoving space. Hence, our analysis was performed twice. In
the case of redshift space, we added an apparent displacement corresponding
to the peculiar velocity of the particles along one axis divided by the
Hubble constant. The comparison of results from real and redshift space
allows us to establish the influence of the velocity dispersion on the
final estimates.

Fig.  1 shows a wedge diagram of the simulated matter distribution in
redshift space for the $\Lambda$CDM model at redshifts $z=0$.  To each
particle we assigned a luminosity chosen at random from a Schechter
luminosity function typical for the LCRS galaxies (Lin et al.  1996).  We
projected the particles into a wedge of angular extension $80\deg\times
1.5\deg$, and we kept simulated galaxies in the magnitude range $14 < m <
18$, about the range of the LCRS.  The mock sample contains about 7500
galaxies.  The figure illustrates that the overdense regions form almost a
percolating system, with huge structured systems between radial velocities
of $(10 000 - 40 000)$ km/s.  A sparse filamentary matter distribution
occupies the low-density regions.

\section{Core-sampling, Cluster analysis, and Minimal Spanning
Tree techniques}

In this paper we are interested in the investigation of large scale
structure, and specific methods are to be used in order to characterize the
simulated matter distribution.  These methods are not so popular as, for
example, the correlation analysis, but they are well defined and allow us
to characterize the matter distribution on large scales comparable with
observations and with the scales predicted by theoretical considerations.

\subsection{Core-sampling approach}

The core-sampling method proposed by Buryak et al.  (1994) was described in
detail in LCRS1 and tested with Voronoi tesselations by Doroshkevich,
Gottl\"ober, \& Madsen (1997).  It allows us to discriminate the
filamentary and the sheet-like structure elements and to find two
quantitative characteristics of the structure, namely, the surface density
of filaments, $\sigma_f$, that is the mean number of filaments crossing a
randomly oriented unit area (i.e.\ 1 $h^{-2}$Mpc$^2$), and the linear
density of sheets, $\sigma_s$, that is the number of sheets crossing the
unit length (i.e.\ 1 $h^{-1}$Mpc) of a random straight line.  These
parameters are equivalent to the mean separation between sheet-like
structure elements, $D_s$, and filaments, $D_f$:  $$D_s = 1/\sigma_s,
\qquad D_f =\sigma_f^{-1/2}, \eqno(3.1)$$ i.e.  these lengths represent the
mean free path between sheet-like and filamentary structure elements.

The core-sampling method also allows us to determine the masses and
velocities of structure elements that intersect a sampling core, i.e.
it provides the surface mass density and dynamical characteristics of
structure elements. These parameters are used for the comparison with
theoretical expectations.

\subsection{Minimal Spanning Tree technique}

The MST is a {\it unique network} associated with a given point sample and
connects all points of the sample to a {\it tree} in a special and unique
manner.  Some definitions and capabilities of this approach are described
by Barrow, Bhavsar, \& Sonoda (1985) and van~de~Weygaert (1991).  Here we
will restrict our investigation to the analysis of the {\it frequency
distribution of the MST edge lengths} $W_{MST}(l)$ (the FDMST method).
The potential of the MST approach is not exhausted by this application.
It allows us to characterize, in particular, the morphology of structure
elements and the typical size of the structure network.

At large distances any correlations between the particle positions are
small, and it can be expected, that the edge length distribution
$W_{MST}(l)$ is similar to that of a Poisson distribution.  But for
filaments this distribution will be dominated by a Poisson distribution
with 1-dimensional support (1D), whereas for sheet-like elements a nearly
2-dimensional (2D) random point distribution is typical.  This means that
the function $W_{MST}(l)$ can be used to characterize statistically the
dominant point distribution in the sample.  To do this, the FDMST can be
fitted to the six-parameter function
$$W_{MST}(x)=-W_0~{dF_f\over dx}~e^{-F_f(x)},\quad x=l/\langle l_{MST}\rangle,
\eqno(3.2)$$
$$F_f(x)=(\beta_1x^{p_1}+\beta_2 x^{p_2})^{p_3},$$
$$ p(x) = {x\over F_f}{dF_f\over dx}=
  p_3{p_1\beta_1 x^{p_1}+p_2\beta_2 x^{p_2}\over
       \beta_1 x^{p_1}+ \beta_2 x^{p_2}},$$
where $l$ and $\langle l_{MST}\rangle$ are the edge lengths and the mean edge
length of
the tree, and $W_0$ provides the normalization of the FDMST.  The function
$F$ represents a power law both for small and large $x$, but it allows a
continuous variation of the power index $p(x)$ with the edge lengths $x$.

Here we are mainly interested in the power index $p(x)$ for larger $x$
that characterizes the underlying geometry of the point distribution on
large scales.  A Poisson point distribution with 1D and 2D support is
characterized by the power indices $p = 1 \,\mbox{and}\, ~2$,
respectively.  Therefore, the asymptote of the function $p(x)$ at large
$x$ characterizes the geometry of the structure elements, and it can be
compared with similar parameters recently found for the observed galaxy
distribution (LCRS2).  This approach was tested with the simulations of
1D, 2D and 3D Poissonian-like point distributions.

\subsection{Cluster analysis and variations of the number of clusters
vs. linking length -- the NCLL method}

The standard cluster analysis (friend-of-friends method) is used widely in
numerical simulations and is well known (see, e.g., Sahni \& Coles 1995).
Here we employ it, first of all, in order to define the structure elements
for a more detailed investigation of their properties.  The cluster
analysis can also be used for the description of the matter distribution
on large scales.  The function $W_{MST}(l)$ is closely connected with the
number of clusters $N^{(t)}_{cl}(r_{link})$ because
$$N^{(t)}_{cl}(r_{link}) = N_p \int_{r_{link}}^\infty W_{MST}(l)dl,
\eqno(3.3)$$
where $N_p$ is the number of points in the sample under investigation.
Therefore, the FDMST and NCLL approaches are similar in many respects.

We use a five-parameter fit of the cluster number vs. linking length
relations, $N_{cl}^{(t)}(r_{link})$:
$$N_{cl}(b)~=~ N_{p}~e^{-F_f(b)},\quad b = \left({4\pi\over 3} n_{p}
r_{link}^3\right)^{1/3}.\eqno(3.4)$$
Here $F_f(b)$ is defined by (3.2), $b$ is the dimensionless linking
length, $\beta_1, \beta_2, p_1, p_2 ~\mbox{and}~ p_3$ are dimensionless
fit parameters, and $n_p$ is the 3D number density of points.

Here we are mainly interested in the variation of the power index $p(b)$
at large $b$.  This method is complementary to the FDMST analysis, and it
allows us to get an independent fit to the power index $p(b)$ at large
$b$.

The NCLL method can also be extended, and, for the more detailed
characteristics of the matter distribution, the variation of single
particles, doublets, triplets and other clusters vs.  linking length can
be considered as well.  In this paper, we consider only the total number
of clusters $N_{cl}^{(t)}(r_{link})$ for the comparison with results
obtained with the FDMST.

\subsection{Inertia tensor method}

The sizes of structure elements can be found with the inertia tensor
method (Vishniac 1986; Babul \& Starkman 1992).  For each
cluster the inertia tensor $I_{ij}$ is expressed as
$$I_{ij} = {5\over N_{mem}} \sum_{N_{mem}}(q_i-q_i^{(0)})(q_j-q_j^{(0)})
\eqno(3.5)$$
where $q_i$ and $q_i^{(0)}$ are the coordinates of the particles and
of the center of mass of the cluster, respectively, $N_{mem}$ is the
number of cluster members.  The conventional normalization $5$ has been
taken to be consistent with a homogeneous ellipsoid.  Hence, the
principal values of the tensor $I_{ij}$, namely the length (diameter),
$L$, the width, $w$, and the thickness, $t$, ($L\geq w\geq t$), give
us objective estimates of the cluster size and of the volume, $V$, and of
the mean overdensity of cluster, $\delta$
$$V={\pi\over 6}L w t,\quad  \delta={N_{mem}\over n_pV} \; .\eqno(3.6)$$
where $N_{mem}$ is a number of point in cluster.

This raw estimate is clearly of limited accuracy, but it is easy to
calculate.  The reliability of this estimate is high for compact regular
clusters, and in general it depends on the cluster shape, in particular on
its lumpiness (see, e.g., Sathyaprakash et al. 1998).  We found from the
simulations, that for large linking lengths, clusters are very lumpy in
its outer regions.  Then the ellipsoidal approximation leads to an
artificial growth of the width and the thickness of the clusters.
However, the cluster diameter $L$ provides a stable characteristic of
the cluster size.

\section{Rich structure elements in simulations}

The cluster analysis shows that rich structure elements (RSE) are usually
represented by rather compact wall-like objects, and our methods give us
more reliable information about their properties, some of which can be
directly connected with the parameters of the cosmological model (see
discussion in Sec.  7).  In contrast, the discrimination and
identification of poor structure elements is always difficult, as in the
observed galaxy distribution they usually represent a filamentary
component in a random network.  This means that the discrimination and
statistical description of such elements is often uncertain as their
shapes are entangled due to many irregular branches.

Because of this, in this paper we give the main attention to the RSE.
Some statistical parameters of the filamentary component have nonetheless
be found.  They are discussed below.

\begin{table*}
\begin{minipage}{180mm}
\caption{Parameters of rich structure elements in comoving (cm) and
redshift (rs) space.}
\label{tbl2}
\begin{tabular}{cccccccccccc} 
model&$z$&$r_{link}$&$b^3$&$b^3_{perc}$&$N_{rse}$&$f_{rse}$&
$\delta_{rse}$&$\sigma_u$&$\sigma_1$&$\sigma_2$&$\sigma_3$\cr
     & &$h^{-1}$Mpc&  &&   & && km/s   & km/s  & km/s& km/s\cr
SCDM-cm        &0&0.75  &0.38&2.22& 1370&0.46 &43.1&670&690&670&590\cr
SCDM-rs        &0&0.75  &0.38&1.95& 1134&0.45 &10.2&610&716&743&703\cr
OCDM-cm        &0&0.95  &0.77&1.39&~~474&0.39&~~4.0&416&333&341&343\cr
OCDM-rs        &0&0.95  &0.77&1.45&~~469&0.44&~~3.8&400&325&340&344\cr
$\Lambda$CDM-cm&0&1.~~~~&0.90&1.52&~~752&0.42 &38.1&596&610&582&514\cr
$\Lambda$CDM-rs&0&1.~~~~&0.90&1.30&~~697&0.45 &14.1&548&565&590&530\cr
$\Lambda$CDM-cm&1&1.~~~~&0.90&1.37&~~561&0.24 &12.1&496&510&527&500\cr
$\Lambda$CDM-rs&1&1.~~~~&0.90&1.44&~~599&0.28 &10.1&458&473&512&497\cr
\end{tabular}

$r_{link}$, $b$ are the threshold linking lengths, $b_{perc}$
characterizes (approximate) percolation, $N_{rse}$, $f_{rse}$ and
$\delta_{rse}$ are the number, fraction of accumulated particles and
mean overdensity, $\sigma_u$ is the dispersion of velocity of RSE and
$\sigma_1$,$\sigma_2$ and $\sigma_3$ are the velocity dispersion of
matter within the structure elements along the three principal axes
of their inertia tensor.
\end{minipage}
\end{table*}

\subsection{Identification of structure elements in simulations}

The large size of the computation box allows us to obtain a representative
set of large clusters which can be associated with the observed RSE.  The
clusters were found for different richness thresholds $N_{thr}$, and for
varying linking lengths $r_{link}$.  The linking length is directly
connected with the density threshold bounding structure elements,
$n_{thr}$.  Indeed, no particles of a cluster are separated from the
neighbor by more than the distance $r_{link}$, therefore a lower limit to
the cluster density is
$$n_{thr}\geq n_{p}b^{-3}.\eqno(4.1)$$
where $b$ is given by (3.4).

For more interesting models the parameters $N_{thr}$ and $b$, used for
the identification of structure elements, as well as the number of
identified structure elements, $N_{rse}$, are listed in Table 2.  We also
give the value $b_{perc}$ which corresponds to the linking length when
the largest cluster of the sample accumulates $\sim$$(25 - 30)$\% of the
points.  This is similar to, but not the exact percolation threshold
(Klypin \& Shandarin 1992).  The smooth shape of FDMST (no break or
cutoffs, see the discussion in Sec.  6.6) shows, that in the simulated
matter distribution even for large $b$, a significant fraction of points
is not accumulated by the largest cluster.  Nevertheless, the fast growth
of the largest cluster distorts the cluster properties, therefore smaller
linking lengths must be used to obtain the typical characteristics of
rich structure elements.

A few parameters for the RSE are applied to discriminate cosmological
models, to select the most realistic models for a detailed analysis, and
to find a reasonable range of the threshold for the discrimination
between RSE and LDR.  These parameters, also listed in Table 2, are the
fraction of matter accumulated by such structure elements, $f_{rse}$, the
overdensity $\delta_{rse}$ given by (3.6), the velocity dispersion of
these structure elements, $\sigma_u$, and the velocity dispersion of
matter within the structure elements along the three principal axes of
their inertia tensor, $\sigma_1,~\sigma_2, ~\sigma_3$.  The overdensity,
the velocity dispersion of the clusters, and the inner velocity
dispersion were averaged over all clusters in the sample within the range
of richness under consideration.  Each cluster was weighted by the number
of cluster members, $N_{mem}$, as this provides parameters typical for
the majority of considered points.  These characteristics are sensitive
to $\Omega_m$ and $h$, and to the amplitude and the shape of power
spectrum.  They are found for all simulations.  The overdensity and the
component of the inner velocity dispersion along the axis with the
smallest component of the inertia tensor, $\sigma_3$, are plotted in
Figs.  2 and 3 for SCDM, OCDM and $\Lambda$CDM models.

\begin{figure}
\centering
\epsfxsize=7 cm
\epsfbox{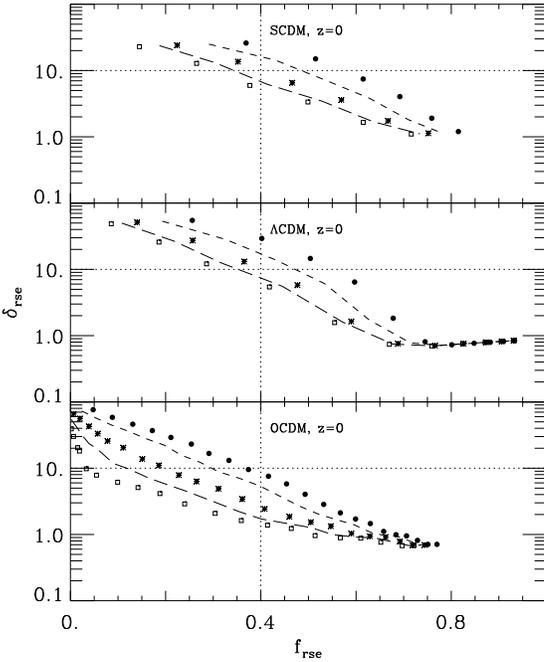}
\vspace{0.5cm}
\caption{Overdensity $\delta_{rse}$ in redshift space vs.
matter fraction concentrated within RSE, $f_{rse}$, for five
richness thresholds: $N_{thr}=100$ (dots), $N_{thr}=200$
(dashed line), $N_{thr}=300$ (stars),
$N_{thr}=500$ (long-dashed line), $N_{thr}=1000$ (open squares).
Dotted lines show the observed parameters of RSE.}
\end{figure}

\begin{figure}
\centering
\epsfxsize=7 cm
\epsfbox{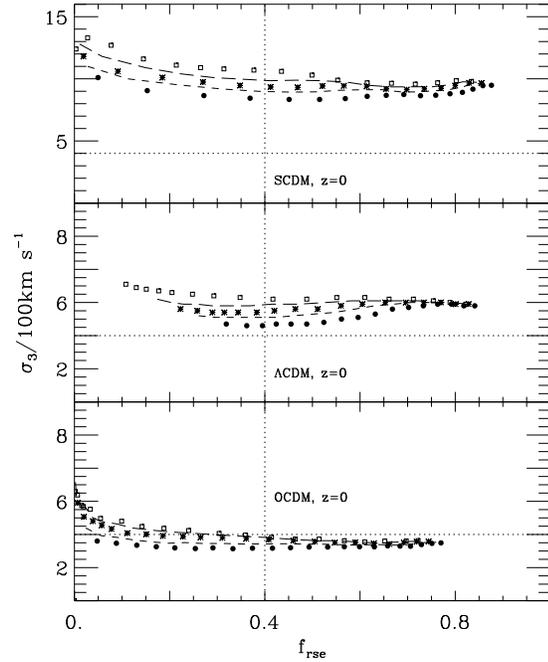}
\vspace{0.5cm}
\caption{Velocity dispersion $\sigma_{3}$ in redshift
space vs. matter fraction concentrated within RSE, $f_{rse}$,
for the same five richness thresholds.
Dotted lines show the observed parameters of RSE.}
\end{figure}

\subsection{Overdensity and velocity dispersions of rich
structure elements}

The simulated characteristics of the RSE must be compared with the observed
parameters.  The velocity dispersion within wall-like structure elements
was roughly estimated by Oort (1983a) to be $\sigma_v^{obs} \sim (350 - 400)$
km/s.  An estimate of the bulk velocity $\sigma_u^{obs} \sim
400$ km/s can be taken from Dekel (1997).  The fraction of galaxies
accumulated by the RSE, $f_{rse}$, and the corresponding overdensity
$\delta_{rse}$, were estimated for the LCRS as $f_{rse} \sim 0.4 - 0.5$
and $\delta_{rse} \sim 10$ (LCRS1 and LCRS2).  The spatial distribution
of RSE can be characterized by the mean separation of RSE along a random
straight line, which was found as $\sim$$(50 - 60)h^{-1}$Mpc in the LCRS.
Using these estimates, we can also find an approximate demarcation between
RSE and LDR.

The analysis shows that the mean velocity of the clusters $\langle
u\rangle$ is negligible in comparison to the velocity dispersion
$\sigma_u$.  This dispersion depends only weakly on the cluster richness,
but it is sensitive to parameters of the cosmological model.  Because of
the vortex-free character of the initial velocity field the velocity $u$
measures the random variation of the gravitational potential over the
cluster and, therefore, the value $\sigma_u$ is approximately proportional
to the amplitude of initial perturbations given by the mean velocity
dispersion, $\sigma_{vel}$, listed in Table 1.  Theoretical considerations
(Demia\'nski \& Doroshkevich 1999b) describe this connection quantitatively
(see also the discussion in Sec. 7).  It can be expected that $\sigma_u$
is slightly smaller in the MDM model where the fraction of hot DM particles
makes the potential distribution more smooth.  For the SCDM and TCDM
models, the dispersions $\sigma_u$ exceed the observed value by about a
factor of 2, and by a factor of about 1.5 for $\Lambda$CDM model.  For the
OCDM and BCDM models, the simulated and observed velocity dispersions are
in general consistent.

The mean overdensity and the inner velocity dispersion, $\sigma_3$, along
the shorter cluster axis are very sensitive both to the cosmological model
and to the subsample of clusters under investigation.  For the most
realistic models, they allow us to estimate a suitable range for the
linking length, $r_{link}$, and the threshold richness of clusters,
$N_{thr}$. For three models the overdensity, $\delta_{rse}$, and the
inner velocity dispersion, $\sigma_3$, are plotted in Figs. 2 and 3 vs.
the matter fraction of the clusters, using five richness thresholds,
$N_{thr}$. In order to compare the parameters of the clusters with
observations, the analysis has been performed in redshift space. The
observed estimates of overdensity, galaxy concentration and velocity
dispersion along the shorter principal axis of RSE, $\sigma_3$, are
plotted in Figs.  2 and 3 as well.

The simulated values of the velocity dispersion $\sigma_3$ exceed the
theoretical expectations (Demia\'nski \& Doroshkevich, 1999b) due to the
wall disruption and the formation of high density clumps.  This is clearly
seen from the isotropy of dispersion listed in Table 2 (we have always
$\sigma_1 \sim \sigma_2 \sim \sigma_3$).  The rate of this disruption
depends on the degree of matter compression and, therefore, on the mean
overdensity of the clusters.

The simulations with $\Omega_m=1$ (SCDM, TCDM and BCDM models) cannot
reproduce the main observed characteristics of the RSE over the range of
considered richness thresholds $100<N_{thr}<1000$ and over a reasonable
range of linking lengths.  For the SCDM and TCDM models, a reasonable
matter concentration is connected with a very large velocity dispersion.
Therefore, even moderate changes in the power spectrum normalization of
these models cannot improve the parameters of the rich clusters.  For the
BCDM model, a reasonable velocity dispersion is accompanied by a very
small matter concentration within the RSE.  For this model a large bias
between the galaxies and the dark matter distribution (a large factor
$\sim 2 - 3$ is required) could, in principle, reduce this disagreement.
The same effect can also be reached by an increase of the amplitude of the
power spectrum by a factor of about two in comparison to the used
normalization.  Thus, already the first step of our analysis shows that
probably the models with $\Omega_m =1$ cannot be considered as realistic.

Models with a smaller matter density $\Omega_m=0.5$ (OCDM) and
$\Omega_m=0.35$ ($\Lambda$CDM) show better results.  For the $\Lambda$CDM
simulation reasonable structure parameters are found for $b^3=0.90$
and $N_{thr}=200$.  Some excess in the velocity dispersion $\sigma_u$ and
in $\sigma_1$, $\sigma_2$, and $\sigma_3$ (a factor of about 1.5) points
to the overevolution of this model.  Better results can be reached by the
variation of the DM composition and/or for smaller values of the
cosmological parameters $h\Omega_m$.  For the simulated OCDM model, all
parameters of RSE at $b^3=0.77$ and $N_{thr}=200$ are consistent to the
observed ones in the range of our precision.  A deficit in the
overdensity ($\delta_{rse} \sim 4$) is within the range of a possible
large scale bias, i.e.  a higher concentration of luminous matter
(galaxies) in rich structure elements in comparison to the concentration
of DM (see Sec.  8).

The considered low density models are most promising for a detailed
investigation.  In these models the COBE normalization is also consistent
with the observed characteristics of rich clusters of galaxy
(cp., e.g. Cole et al. 1997; Bahcall \& Fan 1998).
The curves in Figs.  2 and 3 allow us to establish a rough boundary
between the RSE and LDR, both in terms of the variables $N_{thr}$ and
$r_{link}$ (or $b$), and in terms of physical variables $f_{rse}$ and
$\delta_{rse}=\langle n_{rse}\rangle/\langle n_{p}\rangle$.  Furthermore,
detailed investigation of RSE and LDR allow us to test and to correct
this demarcation.

A significant redshift dependence of the main parameters of RSE is found
for $\Lambda$CDM models at $0\leq z\leq 3$.  Thus, already at $z=1$ a
smaller matter concentration is found, and at $z=3$, the matter fraction
in RSE, $f_{rse} \sim 0.03 - 0.05$, is negligible.  This means that for
these models the RSE are sensitive indicators of the initial amplitude of
perturbations.

\section{Mean separation of filamentary and sheet-like components in
the SCDM, OCDM and $\Lambda$CDM models}

In this Section, properties of simulated structures are examined with the
core-sampling method.  This method allows us to find the mean separations
between filamentary and sheet-like structure elements, respectively.  The
analysis of samples obtained by systematic rejection of sparser
structure elements allows us to reveal the characteristics of typical
structures. These data can be compared with similar results obtained for
the LCRS (cp.  LCRS1). The analysis was performed for the SCDM and OCDM
models at $z=0$, and for the $\Lambda$CDM model at $z=0$ and $z=1$.

For the core-sampling analysis a set of 196 cylinders with a radius of
$1.7h^{-1}$Mpc was prepared both in comoving and in redshift space.  The
mean number of points within the cylindrical cores amounts $\sim$$(400 -
600)$.  The analysis was performed for 16 values of the cylinder radius,
$1.7h^{-1}$Mpc $\geq r_{cyl}\geq 0.7h^{-1}$Mpc.  The separation of
sheet-like elements, $D_s$, and the surface density of the filamentary
component $\sigma_f = D_f^{-2}$, are plotted in Fig.  4 versus the
fraction of matter $f$, remaining after rejection of sparse structure
elements.  It shows the OCDM and $\Lambda$CDM models in redshift space at
$z=0$.  The parameters typical for the reliable structure elements are
marked in Fig.  4 by dotted lines and listed in Table 3 together with
similar parameters obtained for the LCRS (LCRS-1).  Results are found to
be close both in comoving and redshift space, and they coincide with the
parameter range estimated for the LCRS.

\begin{figure}
\centering
\epsfxsize=7.3 cm
\epsfbox{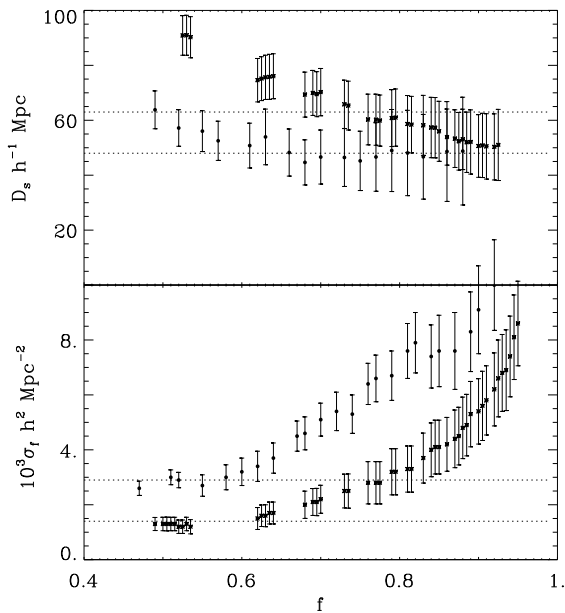}
\vspace{0.5cm}
\caption{Mean separation of the RSE, $D_s$, and the
surface density of filamentary component, $\sigma_f$, vs.
the matter fraction concentrated within the structures, $f$,
in redshift space, and at $z=0$, for the OCDM (dots) and
$\Lambda$CDM (stars) models.
}
\end{figure}

\begin{table}
\caption{Structure parameters with the core-sampling method
in redshift space.}
\label{tbl3}
\begin{tabular}{cccccc} 
 model&z&$N_{thr}$&$D_s$&$D_f$&$D_f^{min}$\cr
   &  &    &$h^{-1}$Mpc&$h^{-1}$Mpc&$h^{-1}$Mpc\cr
LCRS            &0.&--&$55.\pm~~7$&$26.7\pm 3.$&$10.5\pm 1.$\cr
SCDM            &0.&8 &$38.\pm~~9$&$19.2\pm 3.$&$~~7.5\pm 2.$\cr
OCDM            &0.&6 &$48.\pm 11$&$19.2\pm 3.$&$10.0\pm 3.$\cr
$\Lambda$CDM    &0.&4 &$63.\pm 15$&$26.7\pm 5.$&$12.0\pm 4.$\cr
$\Lambda$CDM    &1.&4 &$45.\pm 11$&$23.6\pm 3.$&$~~7.8\pm 2.$\cr
\end{tabular}

The threshold $N_{thr}$ is the minimal richness of the considered
structure elements, $D_s$, $D_f$, and $D_f^{min}$ are the mean separation
of sheet-like elements and rich and poor filaments.
\end{table}

In all cases there is a clear signal from the SLSS component, but in
contrast with results found for the LCRS, $D_s$ increases slowly for
small $f \leq (0.6 - 0.7)$.  This effect is probably caused by the
variation of the covering factor of the sheet-like component, as
discussed by Ramella et al.  (1992) and Buryak et al.  (1994).  This
effect is less prominent for the OCDM model for which the disruption of
RSE measured by the velocity dispersions, $\sigma_1$, $\sigma_2$, and
$\sigma_3$, is also less significant.  The weak variation of $D_s(f)$ and
the quick drop of $\sigma_f$ with $f$ shows that a significant matter
fraction ($\sim 0.4 - 0.5$) is associated with the high dense sheet-like
component that is also consistent with the observational results in the
LCRS (LCRS1).  A population of rich filaments can be identified in Fig.
4, but in contrast with results from the LCRS, it is less representative.
For the SCDM model we also found a strong variation of the number of
particles from core to core.  This is probably caused by the strong
disruption of structure elements.

The parameters $D_f^{min}$ listed in Table 3 correspond to the minimal
separation of filaments found in our analysis at $f \rightarrow 1$.  As it
is common in simulations, the identification of poor filaments is
difficult, and their parameters depend on the rejected background.  It is
well known that for a CDM power spectrum, very small DM pancakes form
early, and the estimates of the minimum pancake size in simulations reflect
the resolution (in our case given by the size of the computation cells).
This means that $D^{min}_f$ listed in Table 3 generally characterizes the
procedure of background rejection rather than properties of poor structure
elements.  This problem was discussed in detail by Doroshkevich et al.
(1997 \& 1998b).  The uncertainties in our estimates of $D_s$ and $D_f$
are larger than those obtained for the LCRS.  This shows that simulated
DM structures are not fully consistent with the geometrical model on which
the core-sampling method is based.

The comparison of structure parameters found for the $\Lambda$CDM models
at $z=0$ and $z=1$ shows the moderate evolution of structure which is
approximately consistent with theoretical expectations (Demia\'nski \&
Doroshkevich 1999b).

\section{Properties of rich structure elements}

The analysis performed in Sec. 4 and 5 shows that, in fact, for a
significant matter fraction $f_{rse}\sim 0.4$, a strong nonlinear matter
compression results in the formation of massive high density RSE. The
existence of such RSE is a very essential feature of the large scale
matter distribution. A more detailed analysis and the statistical
description of RSE is described in this Section.

\subsection{Discrimination of rich structure elements}

The subsamples of RSE were discriminated and analyzed at redshift $z=0$ in
comoving and redshift space for the $\Lambda$CDM, OCDM and SCDM models, and
for estimating the evolution, a similar subsample of RSE was also analyzed
at redshift $z=1$ for the $\Lambda$CDM model.  As described in Sec.  4, the
RSE were identified with rich clusters found for a suitable linking length,
$r_{link}$, and a richness threshold, $N_{mem}\geq N_{thr}$.  The
probability distribution functions discussed below depend on the used
definition of structure element, and in our approach, on these two
parameters.  The employed parameters of the subsamples are listed in Table
4.

In the SCDM model, RSE are defined with linking lengths $b^3 \sim 0.48$,
that corresponds to a threshold density of clusters $n_{thr} \geq 2.1
\langle n_{p}\rangle$, whereas for OCDM and $\Lambda$CDM models
$b^3 \sim 0.9$ provides better results. For these samples, a
significant fraction $f_{rse}\sim 0.4$ of all matter is contained in
massive overdensity clumps with $N_{mem}\geq N_{thr} = 200$.  The chosen
values of $b^3$ are about $2 - 4$ times smaller than the percolation
threshold, $b^3_{perc}$, also listed in Table 2.  This means that the RSE
defined by these thresholds are actually isolated.  Some parameters
discussed below, in particular the comoving sizes, depend on the chosen
$b^3$, and they increase for larger $b^3$ and/or larger $N_{thr}$.

These parameters are the basis for our selection of RSE in simulations.
A more detailed comparison of observed and simulated RSE implies an
identification of `galaxies' in the simulated DM distribution (see Sec.
8).

In the LCRS a similar fraction of galaxies, $f_{gal} \sim 0.4$, is
concentrated in clusters defined with a threshold linking length of
$b^{-3} \sim (1 - 0.5)$, and a mean overdensity in the RSE of $\sim$10
(LCRS2).  This difference between the threshold densities of RSE used in
simulations and in the LCRS is caused in main by the construction of the
LCRS as a set of six slices with angular size $1^\circ.5$ and the
effective thickness $\sim$$(6 - 7)h^{-1}$Mpc only.

The impact of the slice thickness was tested using the mock
catalogues prepared by Cole et al. (1998).
It was found that the random intersections of RSE
with relatively thin slices result in an artificial destruction of
selected RSE. It is seen as the growth of the threshold linking length
required for the selection of RSE and as stronger random variations
of RSE properties with the linking length. Thus, it was found that for
the full mock catalogues about $(40 - 45)$\% of `galaxies' are incorporated
in RSE already at $b^3 \sim 1$ that is close to the values used above.
In contrast, for the slices with angular size $1^\circ.5$ the same
`galaxy' concentration in RSE is reached for $b^3 \sim (1.5 - 2)$
that is comparable with that used for the LCRS (LCRS2). The small
slice thickness depresses also the percolation and restricts the
sizes of the richest RSE in the LCRS. The random mixture of
fields observed with 50 and 112 fibers (in four slices of the LCRS)
amplifies this artificial destruction of selected RSE.

\subsection{Mass functions of rich structure elements}

The mass function of rich structure elements is written as
$$W_m = {N_{mem}\over N_p}N_{rse}(\nu),~~
\nu=N_{mem}/\langle N_{mem}\rangle \eqno(6.1)$$
where $N_{mem}$ and $N_p$ are the numbers of points in the cluster and in
the sample as a whole, $N_{rse}$ is the number of RSE elements for a given
richness $\nu$, and $\langle N_{mem}\rangle$ is the mean number of points
per RSE element.  This function depends on the degree of matter
concentration in RSE, and on the disruption of structures.  This
disruption divides the large structure elements into a system of high
density clumps bridged by low density regions, and it also increases the
fraction of low mass elements.  The rate of disruption depends on the
density contrast and, therefore, it is sensitive to the cosmological
parameters.

\begin{figure}
\centering
\epsfxsize=7.2 cm
\epsfbox{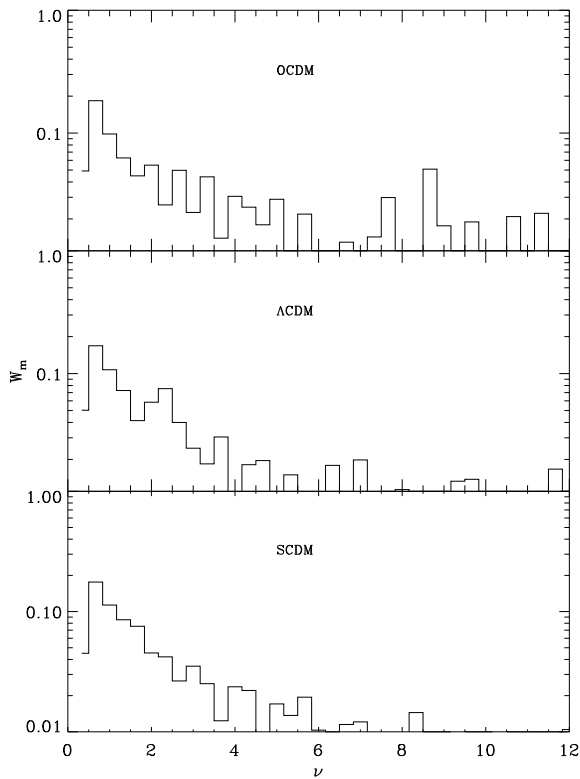}
\vspace{0.5cm}
\caption{Mass function, $W_m$, for the OCDM,~$\Lambda$CDM and
SCDM models in redshift space. Samples and models parameters are listed
in Tables 1, 2 and 4.
}
\end{figure}

The mass functions are plotted in Fig.  5 for the OCDM, $\Lambda$CDM and
SCDM models as determined in redshift space.  The general character of the
functions is very similar in these three models.  Some excess of RSE in
the tail, i.e.  for masses $\nu > (3 - 4)$, are found in the OCDM and
$\Lambda$CDM models.  The threshold richness cuts off the distribution at
small masses, and it influences the value of the mean richness of the RSE,
$\langle N_{mem}\rangle$.  The mean values are listed in Table 4 for all
models under investigation.

These mass functions are in general similar to the observed one in the
LCRS (LCRS2), and at $\nu\leq 6$ they can also be fitted by a simple
exponential law.  In both cases, there are a few huge clusters with mass
$\nu \sim (5 - 10)$ which accumulate $\sim$10\% of points in the simulations.
The rejection of low mass structure elements has a strong influence on the
extent of the mass function.

\subsection{The proper sizes of the rich structure elements}

\begin{table*}
\begin{minipage}{180mm}
\caption{Proper sizes of rich structure elements.}
\label{tbl4}
\begin{tabular}{cccccccccccccc} 
model&z&$r_{lnk}$&$\langle N_{mem}\rangle$&$\langle \lambda\rangle$
&$\sigma_\lambda$&$\langle \omega\rangle$&$\sigma_\omega$&
$\langle \theta\rangle$&$\sigma_\theta$&$\langle D_{prw}\rangle$
&$\sigma_{prw}$&$\langle D_{sep}\rangle$&
$\sigma_{sep}$\cr
LCRS~~~          &0&     &   & 25.3&6.0&12.3
&3.1&5.7&1.3&26.4&1.4&$\sim$38.&$\sim$28.\cr
SCDM-cm          &0&0.75 &731&~~9.2&4.0&~~4.3&1.4&2.7&0.6&16.4&0.6&--  &--\cr
SCDM-rs          &0&0.75 &870& 22.8&6.6&~~7.0&2.6&3.6&0.9&16.0&1.6&--  &--\cr
OCDM-cm          &0&0.95 &907& 24.2&7.8& 12.1&4.4&6.0&1.9&17.3&1.7&40.3&30.3\cr
OCDM-rs          &0&0.95 &998& 24.4&6.8& 12.5&3.2&7.2&1.7&18.3&1.6&37.6&28.3\cr
$\Lambda$CDM-cm  &0&1.0  &801& 14.2&6.2&~~6.2&2.1&3.6&0.9&22.2&1.3&68.4&63.1\cr
$\Lambda$CDM-rs  &0&1.0  &919& 21.8&5.5&~~9.9&2.9&5.0&1.4&23.2&1.2&64.9&57.1\cr
$\Lambda$CDM-cm  &1&1.0  &588& 19.6&6.1&~~7.8&1.8&4.6&1.3&21.4&1.2&79.5&64.0\cr
$\Lambda$CDM-rs  &1&1.0  &641& 19.7&4.4&~~9.1&2.1&4.8&1.1&21.1&0.9&76.3&62.2\cr
\end{tabular}

All mean sizes and dispersions are given by (6.2), (6.4) and Sec. 6.5 and
are measured in $h^{-1}$Mpc.
\end{minipage}
\end{table*}

\begin{figure}
\centering
\epsfxsize=7.15 cm
\epsfbox{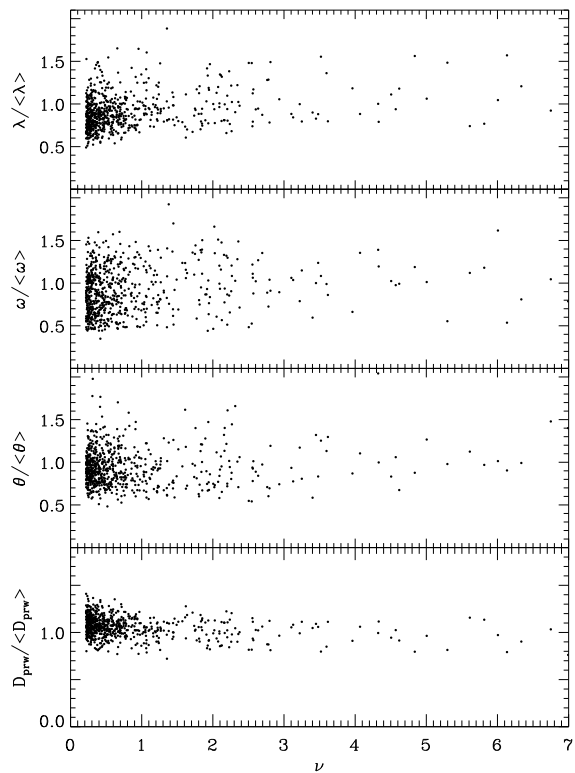}
\vspace{0.5cm}
\caption{The proper sizes of the RSE,
$\lambda/\langle\lambda\rangle, \omega/\langle\omega\rangle,$ and
$\theta/\langle\theta\rangle$, and the size of `proto walls',
$D_{prw}/\langle D_{prw}\rangle$, vs. the mass (richness) of the
element, $\nu = N_{mem}/\langle N_{mem}\rangle$ for the $\Lambda$CDM
model in redshift space at $z=0$.}
\end{figure}

The proper sizes of rich structure elements are found with the inertia
tensor method, applied to the subsample of RSE.  All proper sizes depend
on the mass of the RSE.  The scaling can be approximated by
$$L=\nu^{1/3}\lambda,\quad w=\nu^{1/3}\omega,\quad
t=\nu^{1/3}\theta,\eqno(6.2)$$
$$\sim 0.2\leq\nu=N_{mem}/\langle N_{mem}\rangle\leq 7$$
The mass-averaged length $\langle \lambda\rangle $, width $\langle
\omega\rangle$, and thickness $\langle \theta\rangle$ of clusters are
listed in Table 4 together with the corresponding dispersions.  Scatter
plots of these values vs.  the richness of RSE $\nu$ for the $\Lambda$CDM
model are shown in Fig.  6.  The distributions of proper sizes $\lambda$,
$\omega$, and $\theta$, are similar to Gaussian distributions with mean
values and dispersions listed in Table 4.

The scaling (6.2) describes well the mass dependence of the proper sizes
which results from the relatively regular shape of RSE and the moderate
influence of boundary effects.  In the LCRS a similar scaling is found for
the two larger sizes, whereas the small size is weakly dependent on the
richness (LCRS2).  This is probably caused by the special construction of
the LCRS as a set of thin slices.  The strong richness dependence found
for the Durham/UKST redshift survey (Doroshkevich et al.  1999) could be
caused by the relatively small size of the survey.

Results listed in Table 4 show that for the OCDM and $\Lambda$CDM models,
all proper sizes are found to be close (in the range of the dispersion) to
the sizes observed in the LCRS.  Moderate variations of the threshold
linking lengths do not change the mean characteristics of the RSE, but the
sizes of the largest structure elements are sensitive to such variations.

For the SCDM model, all mean proper sizes in redshift space exceed the
sizes found in comoving space.  These differences become smaller for the
$\Lambda$CDM and the OCDM models, and they decrease with increasing
redshift as it is shown by the comparison of the $\Lambda$CDM model at $z
= 0$ and $z = 1$.  It is explained, in part, by the well known influence
of the velocity dispersion (`finger of God effect').  But this effect also
depresses the small scale clustering and, apparently, partly cancels the
disruption of large structure elements typical for the nonlinear evolution
of pancakes.  This is clearly seen in the growth of the mean length
$\langle \lambda\rangle$, which exceeds any velocity dispersions and,
therefore, is insensitive to the finger of God effect.  In contrast, the
influence of the small scale clustering on the thickness of structure
elements is not so strong, and its growth in redshift space is probably
caused by the direct influence of small scale velocities.  Hence, we can
take the value $\langle L\rangle$ in redshift space as a genuine diameter
of the RSE, and the value $\langle t\rangle$ from comoving space as real
thickness.  For the middle size, $\langle w\rangle$, both effects can be
important.

\subsection{The `proto-size' of the rich structure elements}

The measured parameters of RSE allow us also to estimate the volume which
initially contained all matter in RSE, and the degree of compression
connected with its formation.  A simple model was used for this estimate
in LCRS2.  As the lengths of the RSE exceed the two other sizes, we can
neglect the matter compression along this axis, and we can consider the
RSE formation as a 2D matter infall to the gravitational well.  In this
case we can obtain a simple estimate using the mass conservation:
$$\langle n_{rse} t w\rangle \approx
  \langle n_{p}\rangle \langle t_0 w_0\rangle \approx
  \langle n_{p}\rangle L_{prw}^2
\eqno(6.3)$$
$$L_{prw}^2 = {6 N_{mem} \over \pi L\langle n_{p}\rangle},
  \quad L_{prw}\leq L$$
where $n_{rse}$ and $\langle n_{p}\rangle$ are the comoving number density
of particles within the RSE and in the sample as a whole, respectively,
$t$ and $w$ are the thickness and width of the RSE, and $t_0$,
$w_0$ and $L_{prw} = \sqrt{w_0 t_0}$ are the typical sizes of the
`proto-structures', defined in the initially almost homogeneous matter
distribution. The mass dependence can be described similarly to Eq. (6.2):
$$L_{prw} = \nu^{1/3}D_{prw}\eqno(6.4)$$

The distributions of $D_{prw}$ are also similar to Gaussian distributions
with the mean values, $\langle D_{prw}\rangle$, and dispersions,
$\sigma_{prw}$, listed in Table 4. The comparison of $\omega$ and
$\theta$ with $D_{prw}$ shows that at $z=0$ the RSE formation can be
roughly described as an asymmetric matter compression by a factor of
$\sim$2 along the middle axis ($w$), and a factor of $\sim$$(4 - 5)$ along
the shorter axis ($t$).  At $z = 1$ the corresponding compression factors
are $\sim$1.5 times smaller.  For the high density clumps at redshift
$z=3$ in the $\Lambda$CDM model the size of the `proto-structures' is
$$L_{prw}(z=3)\approx (18\pm 3.5)h^{-1} {\rm Mpc}.\eqno(6.5)$$

\begin{figure}
\centering
\epsfxsize=7 cm
\epsfbox{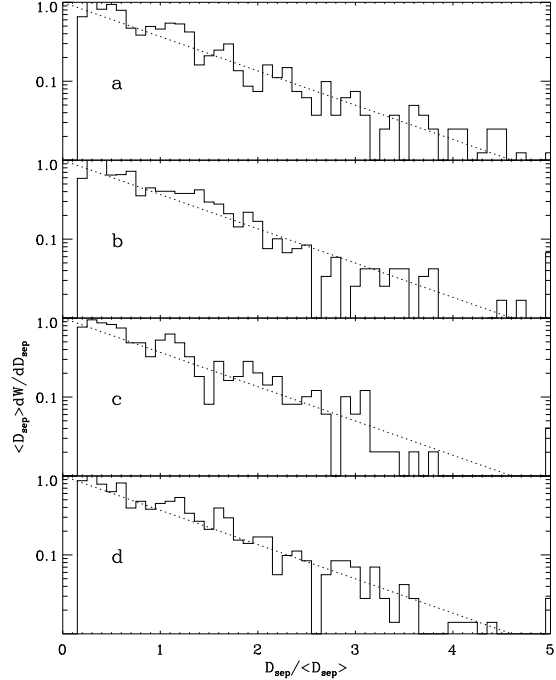}
\vspace{0.5cm}
\caption{The frequency distribution $\langle D_{sep}\rangle dW/dD_{sep}$
of separation of the RSE in comoving (panels a and c) and redshift
(panels b and d) space at z=0 (panels a and b) and z=1
(panels c and d). The exponential fits are shown by dotted lines.}
\end{figure}

The values $D_{prw}$ are plotted in Fig. 6 for given samples of RSE
of the $\Lambda$CDM model.  The parameters are very stable, in particular,
they depend only weakly on the sample under investigation, and on the
linking length, since usually the growth of the cluster sizes is
accompanied by a drop in the number density.  But they are sensitive to
the possible bias between the spatial distribution of DM and galaxies.

\subsection{The mean separation of the structure elements}

The mean separation of RSE can be found, applying a simple version of the
core-sampling method for subsamples of RSE.  A sample of 250 rectangular
cores with sides $10h^{-1}$Mpc$\times 10h^{-1}$Mpc, oriented along one
axis (the axis where the redshift distortions are added) and containing
all particles of RSE, was prepared for the $\Lambda$CDM and OCDM models
both in comoving and in redshift space for $z = 0$ and for the
$\Lambda$CDM model at $z = 1$.  All particles are projected on the axis of
the cores, and they are collected in clusters with the linking length,
$r_{lnk}$, used for the RSE preparation (Table 4). The mean
separation between these clusters (the `mean free-path' between the RSE),
$\langle D_{sep}\rangle$, and the dispersions, $\sigma_{sep}$, are listed
in Table 4.  The large dispersion -- more than 50\% of mean value -- is
typical for an exponential Poisson-like distribution.  The frequency
distribution of cluster separations is plotted in Fig.7 together with the
best exponential fit.  The mean separation of the RSE depends weakly on
the redshift, and it is consistent with the mean separation of wall-like
elements found with the core-sampling approach in Sec.  5.  For the OCDM
model the numerical estimates are consistent to that found in the LCRS
(cp.  LCRS2), but for the $\Lambda$CDM model they exceed the observed
values by a factor of $ \sim$1.5.

The mean separation of RSE (or the 1D number density) allows us to obtain
an independent estimate of the relative richness of RSE for a given
subsample, i.e.  for some linking length $b$ and threshold $N_{thr}$, or a
certain matter fraction in RSE, $f_{rse}$.  The difference found above for
the $\Lambda$CDM model indicates that in this case the same fraction of
particles is concentrated in a smaller number of RSE.  To eliminate this
difference, a threshold richness $N_{thr} \sim 150$ could be used, which
has small effects on the other parameters of RSE.

The mean separations between structure elements of the filamentary
component and their dispersions, $\langle D_f\rangle$ and $\sigma_D$,
can be found through a similar analysis of subsamples prepared
by removing all RSE from the full sample. We obtain $\langle D_f\rangle
\sim 9h^{-1}$Mpc, $\sigma_D\sim 6.5h^{-1}$Mpc, for the OCDM model,
and $\langle D_f\rangle \sim 14h^{-1}$Mpc, $\sigma_D\sim
10.5h^{-1}$Mpc, for the $\Lambda$CDM model.  This data agree well with
the value $D_f^{min}$ listed in Table 3 for the LCRS.

\subsection{Inner structure of RSE and LDR}

The point distribution in a sample can be characterized with the NCLL and
the MST techniques as described in Sec.  3.  These methods allow us to
discriminate between the dominance of filamentary and sheet-like structures
in a point sample, and to characterize the point distribution within
separate structure elements.  In this sense, the methods are complementary
to the core-sampling approach.  Thus, for the filamentary component, power
indices $p_t \,\mbox{and}\, p_{MST}\sim 1$ can be expected, whereas for the
sheet-like component, the appearance of power indices $p_t \,\mbox{and}\,
p_{MST}\sim 2$ seems to be more typical.

Here we apply these methods to the $\Lambda$CDM model.  We analyze
separately the full sample, the RSE, and the LDR, the latter are obtained
by removal of RSE from the full sample.  The main results are presented
in Figs.  8 and 9, and they are collected in Table 5 where $\langle
l_{MST}\rangle$ is the mean edge length of the MST and
$$b_{MST}=({4\pi\over3}\langle n_p\rangle)^{1/3}\langle
l_{MST}\rangle .\eqno(6.6)$$
Here $\langle n_p\rangle$ is the mean number density in the sample under
consideration.

\begin{table}
\caption{Fit parameters for the FDMST and for the cluster distribution,
$p_t$, for full samples, RSE and LDR for the $\Lambda$CDM model.}
\label{tbl5}
\begin{tabular}{ccccc} 
 sample&$\langle l_{MST}\rangle$& $b_{MST}$&$p_{MST}$&$p_{t}$\cr
       &$h^{-1}$Mpc             &          &         &       \cr
\multicolumn{5}{c}{z = 0, comoving space}\cr
a  TOT & 0.72& 0.72&$0.60 \pm  0.02$&$0.59\pm 0.03$\cr
b  RSE & 0.30& 0.62&$1.11 \pm  0.02$&$1.02\pm 0.05$\cr
c  LDR & 1.00& 0.80&$0.83 \pm  0.02$&$0.99\pm 0.03$\cr
\multicolumn{5}{c}{z = 0, redshift space}\cr
d  TOT & 0.78& 0.78&$0.91 \pm  0.03$&$0.85\pm 0.04$\cr
e  RSE & 0.42& 0.83&$1.60 \pm  0.04$&$1.83\pm 0.03$\cr
f  LDR & 1.00& 0.80&$1.11 \pm  0.03$&$1.11\pm 0.02$\cr
\multicolumn{5}{c}{z = 1, comoving space}\cr
   TOT & 0.90& 0.90&$0.76 \pm  0.03$&$0.97\pm 0.03$\cr
   RSE & 0.35& 0.69&$1.0~~\pm  0.03$&$1.30\pm 0.05$\cr
   LDR & 1.00& 0.80&$0.87 \pm  0.03$&$1.10\pm 0.02$\cr
\multicolumn{5}{c}{z = 1, redshift space}\cr
   TOT & 0.91& 0.91&$0.97 \pm  0.02$&$1.02\pm 0.02$\cr
   RSE & 0.45& 0.89&$1.4~~\pm  0.05$&$1.93\pm 0.02$\cr
   LDR & 1.00& 0.80&$1.1~~\pm  0.03$&$1.20\pm 0.03$\cr
\end{tabular}
\end{table}

\begin{figure}
\centering
\epsfxsize=6.3 cm
\epsfbox{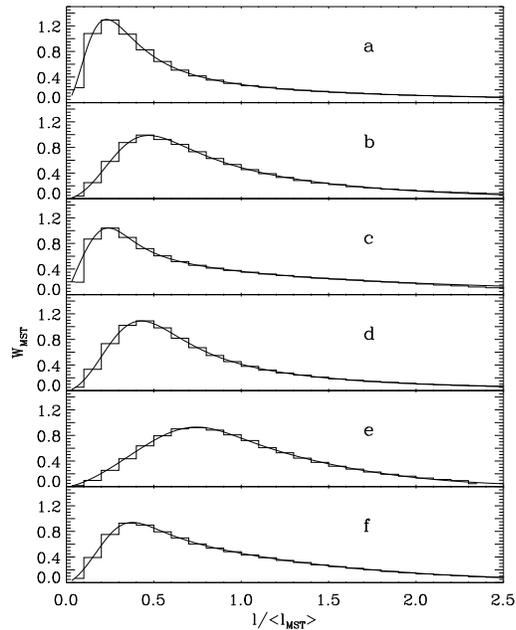}
\vspace{0.5cm}
\caption{The frequency distribution of edges of MST for the $\Lambda$CDM
model.
Sample are marked and sample parameters are listed in Table 5. The fit
of FDMST with relation (3.1) is shown by solid line.
}
\vspace{0.3cm}
\end{figure}

The FDMST are plotted in Fig.  8 for the full sample and for the RSE and
LDR, both in comoving and in redshift spaces, and at $z=0$.  The variation
of the power indices vs.  the edge lengths are plotted in Fig.  9.  The
left-hand side of the FDMST describes the matter condensation within high
density clumps that form the inner structure of filaments and walls.  It is
similar for all subsamples.  The right-hand side of the FDMST characterizes
the relative positions of these clumps and other particles of the
subsamples.

These figures demonstrate that in the comoving space for all samples,
$p_{MST}\sim 1$ for edge lengths $l/\langle l_{MST}\rangle \geq 0.5 -
1$.  The NCLL method confirms that $p_t\sim 1$ is reached for $b\geq
0.5$, this is about 1.6 times smaller than $b\sim 0.8$ used for the RSE
discrimination, and $\sim$2.5 times smaller than the value $b_{perc}$
listed in Table 2.  These values emphasize the joint character of the point
distribution both within RSE and LDR which can be interpreted as a
predominantly 1D Poisson distribution typical for filaments.  The mean edge
length of the MST, $\langle l_{MST}\rangle$, in the RSE is $\sim 2 - 3$
times smaller than within LDR but variations of $b_{MST}$ do not exceed
$\sim$20\%. In redshift space the impact of the velocity dispersion
erases the small scale structure of RSE.  Therefore we find an apparent
particle distribution similar to a 2D Poisson distribution.

The characteristics obtained both with the NCLL and MST methods are similar
to each other within the range of statistical uncertainty.  In any case,
independent on the geometrical interpretation, the power indices and
typical scales, that characterize the spatial matter distribution in the
RSE and LDR, are essentially different, and this verifies the accepted
discrimination of these regions. These results agree well with estimates
of the mean overdensity, $\delta_{rse}$, listed in Table 2 and obtained
in other way. They confirm the essential concentration of high density
clumps in RSE.

The results obtained for the LDR are consistent with the dominance of a
filamentary component. For the RSE, the unexpected value of the power
index in the comoving space can be considered as an indirect evidence
in favor of RSE formation from earlier formed filaments. It can also be
traced back to different factors, such as the wall disruption, the
limited resolution of simulations and, therefore, further investigations
are required.

In the LCRS a power index $p_t \sim p_{MST}\sim 1.7$ has been found
for RSE, and $p_t \sim p_{MST}\sim 1$ for LDR and for the total
sample (LCRS2).  This is comparable with our results for the redshift
space listed in Table 5.  The complicated inner structure of RSE is also
seen in the LCRS and, more clearly, in the galaxy distribution within the
Great Wall (Fig.  5 in Ramella, Geller, \& Huchra 1992).

\begin{figure}
\centering
\epsfxsize=6.6 cm
\epsfbox{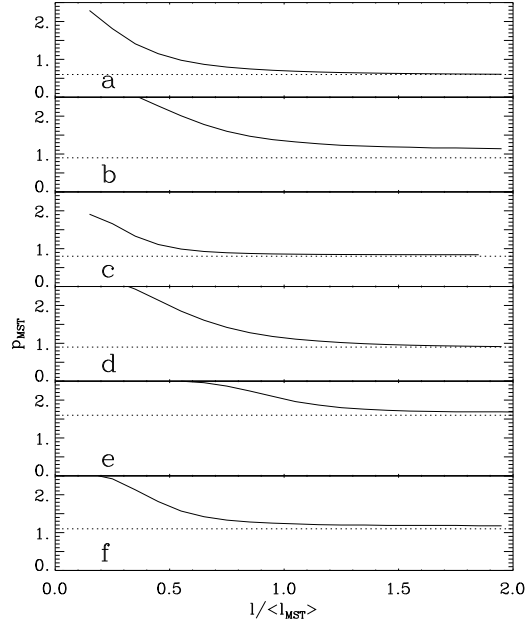}
\vspace{0.5cm}
\caption{Variations of power indices $p_{MST}$ for the same samples
of $\Lambda$CDM model as in Fig. 8. Sample parameters are listed in Table 5.
The fit of FDMST with relation (3.1) is shown by solid line.}
\end{figure}

\section{Characteristics of the expected and simulated DM structure}

The reproduction of the main observed characteristics of the RSE in
simulations with a standard CDM-like power spectrum verifies that the
observed structure was formed during the nonlinear evolution of small
initial perturbations, and, so, the characteristics of structure can be
expressed through the parameters of a suitable initial power spectrum of
Gaussian fluctuations for a specific cosmological model.  Statistical
characteristics of the DM structure based on Zel'dovich nonlinear theory
of gravitational instability (Zel'dovich 1970, 1978; Shandarin \&
Zel'dovich1989) were discussed by Demia\'nski ~\&~ Doroshkevich (1999a, b).
The approximate expressions, derived there, connect some of such
characteristics to the parameters of the cosmological model.  The
comparison of the approximate analytic results and the simulations reveals
both the influence of factors omitted in the theoretical description and
of random factors distorting the simulated structure.

The surface density of RSE and velocity dispersions within RSE, and
the velocity  of structure elements seem to be most interesting. These
values can be found using the simple version of core-sampling described
in Sec. 6.5.  We characterize the mass of each cluster within the
rectangular sampling cores by the surface density of structure elements,
$m_w$, and the velocity dispersion within clusters also along the core,
$\sigma_r$. These values can be found for the RSE. We consider also the
dispersion of 1D velocity of clusters along the core, $\sigma_q$, which
can be found for the full samples, and for the RSE and the filamentary
subsamples of structure elements separately. Some of these characteristics
can be compared with similar 3D characteristics discussed above for RSE
that demonstrates the influence of the sample selection and
the averaging procedure.

\begin{figure}
\centering
\epsfxsize=6.5 cm
\epsfbox{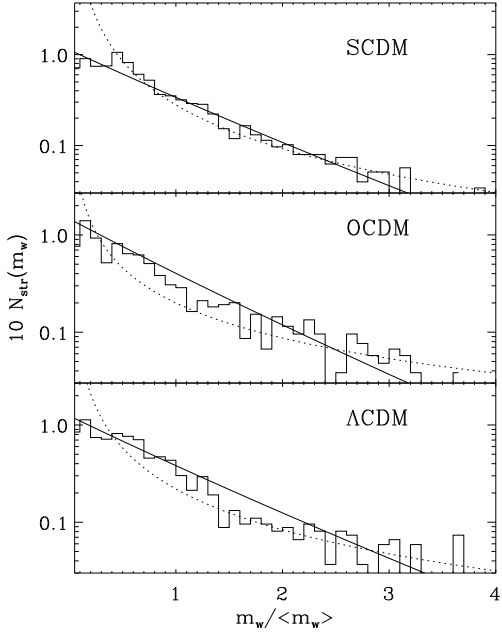}
\vspace{0.75cm}
\caption{The distribution of mass surface density of RSE. The best
fit (7.3) is shown by solid lines, the best fit by power laws is
plotted by dotted lines.
}
\end{figure}

The theoretical parameters are expressed through the typical length scale,
$l_0$, linked to the initial power spectrum, and typical dimensionless
`time', $\tau_0$, linked to the velocity dispersion, $\sigma_{vel}$, and
the parameter $\sigma_8$ listed in Table 1 as:
$$ l_0^{-2}=\int_ {k_{min}}^{k_{max}}k T(k) dk, \quad
l_0\approx {6.6\over \Omega_m h}h^{-1}{\rm Mpc},$$
$$\tau_0 = {\sigma_{vel}\over \sqrt{3} \beta l_0 H_0}, \quad
\beta\approx (0.43+0.57\Omega_m)^{-1},\eqno(7.1)$$
$$\tau_0=0.55\sigma_8(\Omega_mh)^{0.438}\left[1+5.657
(\Omega_mh)^{1.4}\right]^{0.562},$$
where $k$ is the comoving wave number, $k_{min}=2\pi/L_{box},
~k_{max}=k_{min}N_{cell}^{1/3}$, and $T(k)$ is a transfer function.  The
`time' $\tau_0$ characterizes the amplitude of perturbations and the
reached period of structure evolution.  For the SCDM, OCDM and
$\Lambda$CDM simulations we have
$$l_0=13.2,~22,~\&~26.9~h^{-1}{\rm Mpc},$$
$$\tau_0 = \tau_{vel} = 0.43, ~0.27 ~\&~ 0.37\eqno(7.2)$$
$$\tau_0 = \tau_8 = 0.81, ~0.31 ~\&~ 0.46.$$
and differences between values $\tau_{vel}$ \& $\tau_8$ characterize
the sensitivity of various estimates of amplitude to the small scale
matter clustering.

\begin{figure}
\centering
\epsfxsize=6.5 cm
\epsfbox{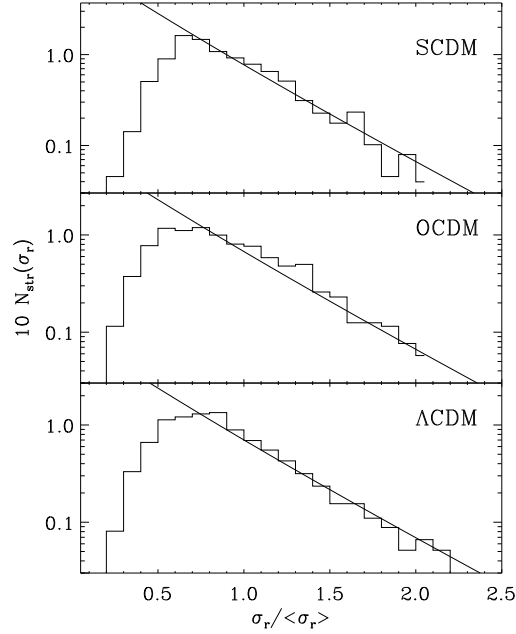}
\vspace{0.75cm}
\caption{The distribution of velocity dispersions $\sigma_r$ for
the RSE. The best fit (7.3) is shown by solid lines.
}
\end{figure}

\begin{table}
\caption{Characteristics of mass and velocity distributions in
comoving space for the full samples, RSE and LDR for the SCDM,
OCDM and $\Lambda$CDM model.}
\label{tbl6}
\begin{tabular}{cccccccc} 
sample&$\sigma_q$&$\sigma_r$&$\mu_w$&$\tau_w$&$\mu_r$&$\tau_r$&$\tau_q$\cr
\multicolumn{7}{c}{SCDM}\cr
TOT & 501& 230& -- & --  &     &     & 0.66\cr
RSE & 442& 492& 1.4& 0.50& 1.29& 0.44& 0.65\cr
LDR & 508& 151& -- & --  & --  &  -- & 0.66\cr
\hline
\multicolumn{7}{c}{OCDM}\cr
TOT & 263&~~98& -- & --  & --  &  -- & 0.30\cr
RSE & 245& 250& 0.8& 0.35& 0.41& 0.25& 0.28\cr
LDR & 277&~~76& -- & --  & --  & --  & 0.32\cr
\hline
\multicolumn{7}{c}{$\Lambda$CDM}\cr
TOT & 360& 155& -- & --  & --  & --  & 0.42\cr
RSE & 351& 412& 1.3& 0.44& 0.63& 0.31& 0.40\cr
LDR & 374& 119& -- & --  & --  & --  & 0.44\cr
\hline
\end{tabular}

the velocity dispersions $\sigma_r$ and $\sigma_q$ are given in km/s.
\end{table}

For Gaussian initial perturbations the distribution of pancake velocities 
along a core is expected also to be Gaussian with a negligible mean 
velocity and a dispersion
$$\sigma_q\approx \sigma_{vel}/3 = H_0l_0\beta\tau_0/\sqrt{3}.
\eqno(7.3)$$
Here a random orientation of pancakes with respect to the sampling cores 
is taken into account.

In this case the distribution of surface density of RSE, $m_w$, can be 
expressed as follows:
$$N_w = {a_w\over \sqrt{x}}e^{-x}erf(\sqrt{x}),
\quad x={b_w m_w\over \langle m_w\rangle},\eqno(7.4)$$
$$\mu_w = {\langle m_w\rangle\over b_w\langle n_p\rangle l_0}
\approx 8(0.5+1/\pi)\tau_0^2= 6.6\tau_0^2 ,$$
where $a_w$ and $b_w$ are fit parameters. The expected distribution
of the filamentary component is described by a more cumbersome relation.

An expression as (7.4) with parameters $a_r$ and $b_r$ and
$x=b_r\sigma_r/\langle \sigma_r\rangle$ describes also the distribution
of the 1D inner velocity dispersion along the core, $\sigma_r$. For
RSE this dispersion can be expressed through $l_0$ and $\tau_0$ as
$$\mu_r = {\sqrt{\langle \sigma_r^2\rangle}\over b_rH_0l_0\beta}\approx
{1\over 2\sqrt{3}}{\langle m_r\rangle\over\langle n_p\rangle l_0}
\approx {3.3\tau_0^2\over \sqrt{3}}\approx \tau_0{3.3\sigma_q\over
H_0l_0\beta}.\eqno(7.5)$$
Using this relation we can measure the mean surface mass density of
RSE $\langle m_r\rangle$ by the velocity dispersion.

These relations allow us to estimate the model parameters $l_0$ and
$\tau_0$ using the measured surface density of RSE $\langle m_w\rangle$,
the velocity dispersion within RSE, $\sigma_r$, and the velocity
dispersions of various populations of structure elements, $\sigma_q$. To
suppress the impact of small scale clustering, the analysis was performed
for the subsample of RSE with the core side $10h^{-1}$Mpc. Because of the
small separation of filaments, for the full sample and subsample of LDR,
discussed in Sec. 6.6, a core side $4h^{-1}$Mpc was used.

For RSE the measured distributions of surface density , $m_w$, and velocity
dispersion, $\sigma_r$, are plotted in Figs. 10 and 11 together
with the best fits (7.4). The measured parameters are listed in Table 6,
the theoretical expectations are given by Eq. (7.2). The distribution
of pancake velocities is well fitted to the Gaussian function with
dispersion $\sigma_q$ listed in Table 6.

These velocity dispersions are smaller by about 20\% than those listed
in Table 2 that demonstrates the influence of averaging procedures in
getting the final parameters.
The values $\tau_0$ averaged over the measurements listed in Table 6, are
$$\langle \tau_0\rangle\approx 0.58\pm 0.09,~~\approx 0.29\pm 0.03,
    ~~\approx 0.40\pm 0.05,\eqno(7.6) $$
for the SCDM, OCDM, and $\Lambda$CDM models.  Differences between these
values and estimates of $\tau_0$ given in (7.2) characterizes the influence
of random factors such as the selection and disruption of structure elements
and the real precision reached.

The measured distributions of surface density of RSE are also well fitted
by power laws with exponents $\kappa \sim 1.7$ that may be caused by the
strong wall disruption. The power law can be reproduced analytically,
assuming a set of clusters with spherically symmetric surface densities
falling off according to a power law $\sigma_{cls}\propto r^{-\gamma}$.
In this case the mass function of clusters in cores is also expressed by a
power law as
$$ N_{cor}dm_w\propto \sigma_{cls}rdr\propto \sigma_{cls}^
{-2/\gamma}d\sigma_{cls},\eqno(7.7)$$
with an exponent $\kappa = 2/\gamma$.  For $\kappa\sim 1.7$ we have
$\gamma \sim 1.2$ that is close to the well known King's law,
$\sigma_{cls}\propto (1+r^2/r_c^2)^{-1/2}, \gamma \sim 1$, widely used to
fit to the density profile of elliptical galaxies.

\section{Analysis of mock catalogues}

For a DM dominated universe the analysis of DM structures is very important
in itself as the feedback of baryonic matter and galaxy formation to the DM
evolution on scales larger than the mean intergalactic separation is
small.  On the other hand, almost all observed characteristics of the large
scale structure are obtained for the galaxy distribution alone, and the
galaxy distribution may be biased in comparison with the distribution of
DM.  Further, all observed galaxy catalogues suffer from selection effects
that influence our cosmological conclusions.

The selection effects are well studied and reproduced in available mock
catalogues (see, e.g., Cole et al. 1998).  A preliminary analysis of these
catalogues reveals, for example, some distortions of observed parameters of
RSE caused by the small angular size of the LCRS (see discussion in Sec.
6). More detailed analysis of these catalogues with the technique
described above is in preparation.  It allows us to find the optimal
strategy of data analysis which suppresses the influence of selection
effects.

\begin{table}
\caption{Parameters of RSE in mock catalogues in redshift space}
\label{tbl7}
\begin{tabular}{ccccccc} 
model&$r_{link}$&$b^3$&$N_{thr}$&$N_{rse}$&$f_{rse}$&$\delta_{rse}$\cr
        &$h^{-1}$Mpc&     &  &   &    &     \cr
mock$_1$&2.1  &0.626&30&830&0.44&~~7.7  \cr
mock$_2$&2.0  &0.586&35&792&0.43&~~9~~  \cr
mock$_3$&1.7  &0.554&50&863&0.45& 11~~  \cr
mock$_4$&1.9  &0.563&70&509&0.45& 14~~  \cr
\end{tabular}
\end{table}

A much more complicated problem is the possible bias between DM and the
galaxy distribution.  The properties of large scale structures are
moderately sensitive to the small scale bias (BBKS; Coles 1993;
for review, Sahni \& Coles 1995), but available observations show that the
spatial distribution of DM and luminous matter can be biased on large
scales as well.  Indeed, while on one hand, in clusters of galaxies the
observed ratio of galaxy and baryonic densities is found to be
$\rho_{gal}/\rho_{gas} \sim 0.2$ (see, e.g., White et al.  1993), on the
other hand, for example within Bo\"ots Void,
$\rho_{gal}/\rho_{gas}\rightarrow 0$ (Weistrop et al.  1992).  The
existence of `invisible' structure elements, which are now seen as gas
clouds responsible for weak Ly-$\alpha$ absorption lines situated far from
galaxies ($\sim$$(5 - 6)h^{-1}$Mpc, Morris et al.  1993; Stocke et al.
1995; Shull et al.  1996) can also be considered as an evidence in favor of
a large scale bias.

Such a large scale bias could be produced by the UV radiation from the
first galaxy population during the reheating of the universe (Dekel \& Silk
1986; Dekel \& Rees 1987).  Quantitative estimates (Demia\'nski \&
Doroshkevich 1997, 1999a, b) show that it can increase the galaxy
concentration within the RSE by about a factor of 1.5 -- 2.  An indirect
evidence in favor of such a bias was found in simulations by Sahni et al.
(1994) as a suppression of structure formation in large regions around the
maxima of gravitational potential, and by Doroshkevich, Fong \&
Makarova (1998) as a difference in the characteristics of spatial
distribution of the rare high peaks identified as `galaxies' and of the
main fraction of structure elements in simulated DM
distributions.  This bias can be essential for the reliability of
discrimination between cosmological models.

\begin{table*}
\begin{minipage}{180mm}
\caption{Proper sizes of RSE in mock catalogues}
\label{tbl8}
\begin{tabular}{ccccccccccccc} 
model&$r_{lnk}$&$\langle N_{mem}\rangle$&$\langle \lambda\rangle$&
$\sigma_\lambda$&$\langle \omega\rangle $&$\sigma_\omega$&
$\langle \theta\rangle $&$\sigma_\theta$&$\langle D_{prw}\rangle$
&$\sigma_{prw}$&$\langle D_{sep}\rangle$&$\sigma_{sep}$\cr
LCRS~~~&& & 25.3&6.0&12.3&3.1&5.7&1.3&26.4&1.4&$\sim$38&$\sim$28\cr
mock$_1$-rd&2.1 & 108~~& 20.1&4.9& 10.6&2.0&6.0&1.4&23.0&1.2&~62&~57 \cr
mock$_2$-rd&2.0 & 119~~& 19.3&4.4& 10.5&2.1&5.9&1.4&23.2&0.9&~53&~57 \cr
mock$_3$-cm&1.9 & 179~~& 18.5&6.2&~~8.4&2.7&4.6&1.3&22.8&1.2&~51&~53 \cr
mock$_3$-rd&1.7 & 174~~& 18.3&4.5&~~9.4&2.2&5.1&1.4&22.9&0.9&~49&~54 \cr
mock$_4$-rd&1.9 & 218~~& 21.0&5.1& 10.7&2.5&5.4&1.8&27.5&0.8&~57&~64 \cr
\end{tabular}

All mean sizes and dispersions are given by (6.2), (6.4) and Sec. 6.5 and
are measured in $h^{-1}$Mpc.
\end{minipage}
\end{table*}

Unfortunately, such large scale bias cannot be simulated yet, since
simulations of the galaxy formation in large boxes with the required
resolution are impossible.  This means that a bias can only be introduced
by hand using simple plausible assumptions.  Some such models as discussed
by Cole et al.  (1998) increase the `galaxy' concentration within RSE.
More detailed tests of these models will be discussed separately.

Here we restrict our consideration to the analysis of simple mock
catalogues prepared for the OCDM model.  With the spatial and mass
resolutions of our simulations ($\sim$$10^{11}M_\odot$), we are compelled
to identify `galaxies' with selected DM particles. Four mock samples with
different clustering
properties were prepared and investigated.  The simplest sample mock$_1$
was constructed with a threshold prescription depending on the local
environmental density within a sphere of $1h^{-1}$Mpc around the
particles.  No `galaxies' are identified with particles with local density
smaller than the mean density, and `galaxy' tracers are selected randomly
from the particles in overdense regions.  The other catalogues use a
smooth probability distribution proportional to the local environmental
density within the same scale.  The constant of proportionality was chosen
to vary the degree of clustering of the mock samples, i.e.  to get an
autocorrelation function of simulated `galaxies' in broad agreement with
data (for the LCRS, cp.  e.g.  Tucker et al.  1997).  The catalogues were
normalized to the mean number density of galaxies, $n_{gal}\sim 2\cdot
10^{-2}h^3$ Mpc$^{-3}$, that is equivalent to the observed galaxy density
with the limiting magnitude of about $M_R = - 18$.

The sample mock$_1$ with the threshold bias shows weakly enhanced
clustering of the mock `galaxies' with respect to the dark matter, the
correlation function in redshift space shows a power law
$(r/r_0)^{-\gamma}$ with a correlation length $r_0=6.5h^{-1}$Mpc and a
slope $\gamma=1.4$.  The correlation length of DM, also in the redshift
space, is $r_0=5 h^{-1}$ Mpc, and the slope is $\gamma=1.3$.

The impact of the local environment allows to vary the clustering and to
obtain mock samples with different properties.  Thus, the weakly clustered
sample mock$_2$ is similar to mock$_1$ (correlation length in redshift
space $r_0=6h^{-1}$Mpc, slope $\gamma=1.4)$ while mock$_3$ is
intermediate ($r_0=7h^{-1}$Mpc, $\gamma=1.5$), and mock$_4$ strongly
clustered ($r_0=8h^{-1}$Mpc, $\gamma=1.6$).  For the sample mock$_2$, we
impose an additional threshold of no `galaxy' identification for lower
than mean density.  We discuss such mock catalogues in our rectangular
slices that allow a direct comparison with the DM catalogue studied above.
It demonstrates directly the influence of bias models.  This is most
important as a first step of the analysis of the influence of bias.  The
selection criteria required for the more detailed comparison with observed
catalogues as the LCRS (cp.  the mock sample for $\Lambda$CDM presented in
Fig.  1) will be imposed as a next step in a separate paper.  The simple
models described above reproduce only some features of the large scale
bias and serve mainly as illustration of the potential of the used
statistics for the bias discrimination.  More realistic models would be
sensitive to the more broad density environment of particles, i.e.  they
are able to follow in more detail the expected interaction of large and
small scale perturbations.  Such models will be studied and discussed
separately.

The first step of our analysis repeats the approach utilized in Sec.  4 to
define the sample of RSE with the required richness and overdensity.  Such
sets of RSE were found in redshift space for all four mock catalogues with
parameters listed in Table 7.  For all mock catalogues the threshold
density of RSE was $\sim$30\% larger than that for the DM catalogue.  For
the mock$_1$ and mock$_2$ catalogues, the small value of $N_{thr}$
requires that the reproduction of a sufficient `galaxy' concentration
within the RSE is also accompanied by a relatively low overdensity of RSE.
Even so, the overdensity is found to be at least two times larger than for
the corresponding DM catalogue (see Table 2), and it is comparable with
the overdensity found for the LCRS.  For mock$_3$ and mock$_4$ catalogues,
the parameters of RSE are similar to each other, and to that found for the
LCRS, and the overdensities reached are about three times larger than for
the DM catalogue.  For all mock catalogues the velocity dispersions are
the same as those found for DM RSE.  These results alone show that the
used models of bias provide an essential excess of `galaxy' concentration
within the RSE and, therefore, they can be considered as a reasonable
basis for the further more detailed investigation.

The proper sizes of rich structure elements are found with the inertia
tensor method, applied to the subsample of RSE.  All proper sizes depend
on the mass of the RSE.  For all mock catalogues the scaling can be
approximated by
$$L=\nu^{0.43}\lambda,\quad w=\nu^{0.46}\omega,$$
$$t=\nu^{0.44}\theta,~L_{prw}=\nu^{0.33}D_{prw}\eqno(8.1)$$
$$\sim 0.3\leq\nu=N_{mem}/\langle N_{mem}\rangle\leq 25.$$
The mass-averaged length $\langle \lambda\rangle$, width $\langle
\omega\rangle $, thickness $\langle \theta\rangle $ and the size of
`proto-wall', $D_{prw}$, are listed in Table 8 together with the
corresponding dispersions.  The stronger scaling found for the mock
catalogues as compared to the DM catalogue is also caused by the biasing
and therefore, this approach can also be used to discriminate bias models.
The main parameters of RSE listed in Table 8 are close to those listed in
Table 4 for the OCDM model in redshift space.  The shape and power indices
found for the FDMST are also close to those listed in Table 5.  As before,
in comoving space the 1D character of the `galaxy' distribution dominates,
i.e.  the exponent $p_{MST}\sim 1$, whereas in redshift space, the `galaxy'
distribution within RSE is similar to a 2D Poissonian distribution with the
exponent $p_{MST} \sim 2$.

These results show that the discussed methods reveal the influence of the
used bias models.  The impact of bias is also essential for the less
massive structure elements, and especially for the matter and `galaxy'
content in `voids'.  These regions will be further investigated with the
MST technique in a separate study.

\section{Summary and Discussion}

In this paper the properties of simulated spatial matter distributions were
studied for five cosmological models with CDM-like power spectra.  The main
parameters of simulations are listed in Table 1.  The simulations were
performed in large boxes in order to reproduce correctly the mutual
interaction of large and small scale perturbations, and to obtain a
representative sample of wall-like RSE.  The broad set of considered
cosmological models allows us to reveal the influence of main cosmological
parameters on the formation and evolution of the wall-like RSE, and to
discriminate between these models.  Our results show that the methods
utilized in this paper are effective, and they yield a description of the
spatial matter distribution on large scales and, in particular, the
characteristics of the RSE.

\subsection{Main results}

The main results of our analysis can be summarized as follows:

\begin{enumerate}

\item{} Simulations performed with the standard COBE normalized CDM-like
power spectrum reproduce well both the wall-like RSE and the filamentary
component of structures in LDR.  Each component accumulates an essential
fraction of matter, and it is equally important for the description of the
joint network structure in the large scale matter distribution.

\item{} The phenomenon of the strong matter concentration within the
wall-like rich structure elements can also be reproduced for suitable
cosmological models.  An essential fraction of DM, $f_{rse}\sim 0.4$, is
compressed nonlinearly on the scales $\sim$$(17 - 25)h^{-1}$Mpc that is
less than the mean separation of these RSE by a factor of $\sim 2 - 3$.

\item{} The RSE are usually disrupted into a system of high-density clumps
that results in the growth of the inner velocity dispersion.  The rate and
the degree of disruption depends on the parameters of the cosmological
models.

\item{} The comparison of observed and simulated parameters of the
wall-like RSE allows us to discriminate between different cosmological
models and to reveal the class of models which can reproduce the main
observed characteristics of the wall-like RSE.  These are the $\Lambda$CDM
model with $\Omega_m h \sim 0.15 - 0.25$ and the OCDM model with
$\Omega_m h \sim 0.25 - 0.35$.  Perhaps, promising results can be also
reached for MDM models with similar parameters.

\item{} A large scale bias between the spatial distribution of DM and
galaxies can increase the galaxy concentration within RSE by a factor of
about 2  which essentially improves the simulated characteristics of
RSE.  The technique used above allows us to reveal reliably the influence
of the biasing and to discriminate between bias models.

\item{} The simulated parameters of DM structure are consistent to the
theoretical expectations.  The main cosmological parameters can be
successfully reconstructed using the measured properties of the large
scale matter distribution.  After correcting for the bias and for
selections effects, these methods can be applied to observed galaxy
catalogues.

\item{} Our results verify also the theoretical expectations with
respect to the epoch of the RSE formation. At $z=1$ the fraction of
matter accumulated by RSE with the chosen richness and overdensity
drops by a factor $\sim$2, and at $z = 3$, it becomes negligible.

\end{enumerate}

The main statistical characteristics of the RSE are listed in Tables 2 --
5 in comparison with the properties found for the observed galaxy
distribution.  These results verify the existence of a wall-like
component with similar characteristics both in observations and
simulations performed for suitable models.  A similar range of
cosmological models was recently separated by Cole et al.  (1997) and by
Bahcall \& Fan (1998) from an comparison of observed and simulated
properties of clusters of galaxies.  The observations of supernovae at
high redshifts (Perlmutter et al.  1998) are more consistent with the
$\Lambda$CDM model with $\Omega_m\sim 0.3$.

Now there is some observational evidence of large matter inhomogeneities
at redshifts $z \sim 0.5 - 1$ and more (Williger at al.  1996;
Cristiani et al.  1996; Quashnock et al.  1996, 1997; Connolly et al.
1996). Our analysis shows that for the considered models, these
structures cannot be as common as at small redshifts.  A more detailed
statistical description of the absorption spectra of quasars is
required to obtain the characteristics of structures at high redshifts.

At small redshifts further progress in investigations of observed large
scale matter distribution is linked with very large galaxy redshift
catalogues as the 2dF redshift survey of 250,000 galaxies (Colless 1998)
and the million galaxy Sloan Digital Sky Survey (Loveday \& Pier 1998).
The available surveys used above for the comparison with simulated
structure parameters (the Durham/UKST Galaxy Redshift Survey, Ratcliffe
et al.  1996; and the Las Campanas Redshift Survey, Shectman et al.
1996) are not so representative and moreover they are limited in use to
specific selection effects (see discussion in Sec.  6.1 \& 6.3).  In
spite of this, now and within the next few years results obtained with
these surveys provide us with the best characteristics of observed large
scale matter distribution.

\subsection{ Methodical remarks}

The study of simulations is now the best way for the understanding of
the large scale matter distribution. The large boxes used for simulations
allow us to obtain a representative description of large scale perturbations
and their mutual interactions with perturbations on smaller scales, as
well as to obtain a representative statistic of RSE.  Both factors are
equally important for the successful reproduction of the matter distribution
observed in large galaxy surveys.

The analysis of six simulations performed by Madsen et al.  (1998) shows
that simulations reproduce the theoretical distributions only on the
scales $\sim$$(0.1 - 0.15) L_{box}$.  This means that realistic simulations
of the observed large scale matter distribution is possible for
$L_{box}\geq (400 - 500)h^{-1}$Mpc, whereas for smaller box sizes random
variations of parameters of large scale structure are expected.

The analysis was performed mainly for rectangular slices with the size
$(500\times 500\times 50) h^{-3}$Mpc$^3$ which accumulate about 10\% of
the particles.  Such an approach allows us to study a broad set of
cosmological models with a reasonable precision and representativity.  To
test the possible impact of the selection used, the analysis was repeated
for the full simulated sample of the $\Lambda$CDM model in comoving
coordinates.  The results are consistent with what was found above, and
the difference is less than 10\%.  The comparison of results obtained for
two $\Lambda$CDM models shows also the difference between the structure
parameters and the velocity dispersions $\sim$10\%.  Variations of
$\sigma_{vel}$ and $\sigma_8$ listed in Table 1 are also $\sim$10\%.
These results characterize the actual precision reached in the
investigation and shows that even for large boxes the main structure
parameters are moderately sensitive to the random realization.

The comparison of presented data with a similar simulation (Cole et al.
1997) performed with higher resolution (P3M code) demonstrates the moderate
dependence of the main simulated structure parameters on these factors.
The properties of high density clumps are sensitive to the resolution that
distorts the FDMST for smaller lengths (in the comoving space). These
distortions however disappear in redshift space. The same factor increases
the simulated velocity dispersion. The main quantitative characteristics
of the RSE are nonetheless sufficiently stable.

\subsection*{Acknowledgments}
We are grateful to our anonymous referee and M.Demia\'nski for the very
useful comments and criticism.  This paper was supported in part by
Denmark's Grundforskningsfond through its support for an establishment of
the Theoretical Astrophysics Center and grant INTAS-93-68.  AGD and VIT
also wish to acknowledge support from the Center of Cosmo-Particle
Physics, Moscow.

\end{document}